\RequirePackage{ifpdf}
\documentclass[aps,twocolumn,nofootinbib,preprintnumbers]{revtex4-1}
\usepackage[T1]{fontenc}
\usepackage{ae,aecompl}
\usepackage{bm}
\usepackage{latexsym}
\usepackage{dcolumn}
\usepackage{amsfonts,amssymb}
\usepackage{graphicx,epsfig}
\usepackage{amsmath,amssymb,multirow,array}

\usepackage{tabularx}
\usepackage{float}
\usepackage{wrapfig}
\usepackage{caption}
\usepackage{subcaption}

\usepackage{hyperref} 
\hypersetup{
	colorlinks=true,	
	linkcolor=blue,		
	citecolor=blue,		
	filecolor=blue,		
	urlcolor=blue		
}

\newcommand{\half}{\frac{1}{2}}

\newcommand{\reff}[1]{(\ref{#1})}

\newcommand{\eq}[1]{Eqn. (\ref{#1})}

\newcommand{\sctn}[1]{Section (\ref{#1})}

\newcommand{\apndx}[1]{Appendix (\ref{#1})}
\newcommand{\apndxIn}[1]{(Appendix \ref{#1})}

\newcommand{\nn}{\nonumber}

\newcommand{\ben}{\begin{eqnarray}\displaystyle}
\newcommand{\een}{\end{eqnarray}}

\newcommand{\be}{\begin{equation}}
\newcommand{\ee}{\end{equation}}

\newcommand{\bea}[1]{\begin{align}#1\end{align}}
\newcommand{\bem}[1]{\begin{multline}#1\end{multline}}
\newcommand{\bee}[1]{\begin{equation}#1\end{equation}}

\newcommand{\dbrk}{\right. \\ \left.}
\newcommand{\tbrk}{\right. \right. \\ \left. \left.}

\newcommand{\dsp}{\displaystyle}

\newcommand{\bc}{\begin{center}}
\newcommand{\ec}{\end{center}}

\newcommand{\grad}{{\rm {\bf \nabla}}}

\newcommand{\Rmnum}[1]{\expandafter\@slowromancap\romannumeral #1@}

\newcommand{\df}{\mathrm{d}}

\renewcommand{\a}{\alpha}	
\renewcommand{\b}{\beta}		
\newcommand{\dow}{\partial}
\newcommand{\e}{\epsilon}

\newcommand{\f}{\phi}

\newcommand{\g}{\gamma}

\renewcommand{\k}{\kappa}	
\renewcommand{\l}{\lambda}	
\renewcommand{\o}{\omega}	
\newcommand{\vp}{\varpi}							

\renewcommand{\r}{\rho}		
\newcommand{\s}{\sigma}
\newcommand{\vs}{\varsigma}
\renewcommand{\t}{\tau}		

\newcommand{\G}{\Gamma}

\renewcommand{\L}{\Lambda}	
\newcommand{\M}{\mho}
\newcommand{\N}{\nabla}
\renewcommand{\O}{\Omega}	

\newcommand{\U}{\Upsilon}

\newcommand{\cE}{\mathcal{E}}

\newcommand{\cK}{\mathcal{K}}

\newcommand{\cO}{\mathcal{O}}

\newcommand{\fa}{\mathfrak{a}}
\newcommand{\fb}{\mathfrak{b}}
\newcommand{\fc}{\mathfrak{c}}
\newcommand{\fd}{\mathfrak{d}}

\newcommand{\ff}{\mathfrak{f}}

\newcommand{\ra}{\rightarrow}
\newcommand{\Ra}{\Rightarrow}

\newcommand{\lB}{\left [}
\newcommand{\rB}{\right ]}
\newcommand{\lb}{\left (}
\newcommand{\rb}{\right )}
\newcommand{\lbr}{\left \{}
\newcommand{\rbr}{\right \}}

\newcommand{\pa}{\partial}

\begin{document}

\title{Entropy current for non-relativistic fluid}
\author{Nabamita Banerjee}
\email[]{E-mail: tpnb@iacs.res.in}
\affiliation{Indian Association for the Cultivation of Science,
  Jadavpur, Kolkata}
\author{Suvankar Dutta}
\email[]{E-mail: suvankar@iiserb.ac.in}
\affiliation{Dept. of Physics, Indian Institute of Science Education
  and Research Bhopal, Bhopal}
\author{Akash Jain}
\email[]{E-mail: ajainphysics@gmail.com}
\affiliation{Dept. of Physics, Indian Institute of Science Education
  and Research Bhopal, Bhopal}
\author{Dibakar Roychowdhury}
\email[]{E-mail: dibakarphys@gmail.com}
\affiliation{Dept. of Physics, Indian Institute of Science Education
  and Research Bhopal, Bhopal}


\begin{abstract}
  We study transport properties of a parity-odd, non-relativistic
  charged fluid in presence of background electric and magnetic
  fields. To obtain stress tensor and charged current for the
  non-relativistic system we start with the most generic relativistic
  fluid, living in one higher dimension and reduce the constituent
  equations along the light-cone direction. We also reduce the
  equation satisfied by the entropy current of the relativistic theory
  and obtain a consistent entropy current for the non-relativistic
  system (we call it ``canonical form'' of the entropy
  current). Demanding that the non-relativistic fluid satisfies the
  second law of thermodynamics we impose constraints on various first
  order transport coefficients. For parity even fluid, this is
  straight forward; it tells us positive definiteness of different
  transport coefficients like viscosity, thermal conductivity,
  electric conductivity etc. However for parity-odd fluid, canonical
  form of the entropy current fails to confirm the second law of
  thermodynamics. Therefore, we need to add two parity-odd vectors to
  the entropy current with arbitrary coefficients. Upon demanding the
  validity of second law, we see that one can fix these two
  coefficients exactly.
 
\end{abstract}


\maketitle

\tableofcontents

\section{Introduction and Summary} \label{intro}

Hydrodynamics is an effective description of nearly equilibrium
interacting many body system. A fluid system is considered to be
continuous, i.e. when we talk about a small volume element (or
`fluid particle') of the fluid, it
still contains a large number of constituent particles
(atoms/molecules).  More specifically, the size of the fluid particle
is much much greater than the mean free path of the system. The
equations of hydrodynamics assume that the fluid is in local
thermodynamic equilibrium at each point in space and time, even though
different thermodynamic quantities like fluid velocity $\vec v(\vec x,
t)$, energy density $\e(\vec x, t)$, pressure $p(\vec x, t)$, fluid
density $\rho(\vec x, t)$, temperature $\t(\vec x, t)$ etc. may
vary. Fluid mechanics applies only when the length scales of variation
of thermodynamic variables are large compared to equilibration length
scale of the fluid, namely the mean free path \cite{landau}.

Although hydrodynamics is an old and well-studied subject in physics,
recently there has been much interest and progress in relativistic,
charged, viscous fluid in presence of some global anomaly. The first
evidence of relativistic anomalous fluids was observed in
\cite{Banerjee:2008th,Erdmenger:2008rm}, in the context of the
$AdS/CFT$ correspondence. In these papers the authors found a new term
(and hence a new transport coefficient) in the charge current in
presence of a Chern-Simons term in the bulk Lagrangian. Later, in
\cite{Son:2009tf} it has been shown that this kind of term in charge
current is not only allowed by the symmetry but is required by
triangle anomalies and the second law of thermodynamics. Demanding the
positivity of local entropy current they showed that the coefficient
of this new term is also fixed in terms of anomaly coefficient of the
theory. In general, the second law of thermodynamics (or equivalently,
positivity of local entropy current) imposes restrictions on different
transport coefficients. Interested readers can look at
\cite{Loganayagam:2008is,Bhattacharyya:2014bha,Bhattacharyya:2013lha,
  Bhattacharyya:2012nq, Banerjee:2012iz, Jensen:2011xb,
  Jensen:2012jh, Banerjee:2012cr}.

This was about relativistic fluids. Much attention has not been paid in
parity-odd charged non-relativistic fluid in presence of background
electromagnetic fields. In \cite{Rangamani:2008gi} an attempt has been
made to study non-relativistic neutral fluids as a consistent
light-cone reduction of relativistic fluid systems. In
\cite{Brattan:2010bw}, the idea has been extended to charged fluid
without any background field. However, main focus of these two papers
was to construct the holographic duals of long wave length
fluctuations (hydrodynamic limit) of conformal non-relativistic field
theories. They also computed few transport coefficients (for example
thermal conductivity) holographically. 

Recently in \cite{Kaminski:2013gca}, non-relativistic, parity-odd,
first order charged fluid in 2 dimensions has been discussed. The
authors started with a $2+1$ dimensional relativistic, parity-odd
charged fluid in presence of background gauge fields and took suitable
non-relativistic limit of the system. They identified the
parity-violating contributions to the non-relativistic constitutive
relations, which include the Hall current flowing perpendicular to the
temperature gradient, the Hall viscosity and the Leduc-Righi energy
current\footnote{See also \cite{Lucas:2014sia}.}. However, in these
papers, much attention has not been paid to second law of
thermodynamics. It is well known that the local second law of
thermodynamics imposes several constraints on the flow and the
transport coefficients. For example, the local second law demands that
shear, bulk viscosity coefficients and thermal conductivity have to be
positive definite for a non-relativistic fluid \cite{landau}. This
particular issue, for relativistic fluid, has been discussed vividly
in various papers mentioned above. However, the constraints from
second law of thermodynamics are not yet well understood for a general
class of non-relativistic fluids (charged fluid in presence of
background fields). Goal of this paper is to write down a local
entropy current, which satisfies second law of thermodynamics, for a
non-relativistic charged fluid in presence of arbitrary background
electromagnetic field and to study the constraints on the transport
coefficients.

One way to obtain the energy momentum tensor, charge current and
entropy current for a non-relativistic theory is to reduce the
constitutive equations and entropy current equation of a relativistic
fluid over a light-cone direction. We start with a $(3+1)$-dimensional
relativistic charged fluid with global $U(1)^3$ anomaly and write down
their non-relativistic counterparts. We follow the light-cone
reduction technique, which was introduced in \cite{Rangamani:2008gi},
and shall as well review it in our paper. Let us conclude this section
with an outline and summary of our main results.\\

\noindent
{\bf Outline and Summary :}

\begin{itemize}

\item 
  In \sctn{sec:relflu} we consider first order, relativistic, charged
  anomalous fluid in $(3+1)$ dimensions in presence of background
  electromagnetic fields. We briefly review the work of
  \cite{Son:2009tf} and show how parity-odd transport coefficients are
  related to the coefficient of anomaly from the positivity of
  divergence of local entropy current.

\item
  In \sctn{sec:non-rel-fluid} we highlight the important equations of
  non-relativistic hydrodynamic system.

\item
  In \sctn{sec:lcr} we consistently obtain the energy current density
  ($j^i$), stress tensor ($t^{ij}$) and charge current density
  $(j_I^i)$ of non-relativistic fluid by Light Cone Reduction
  (LCR). We also find different non-relativistic fluid variables
  (thermodynamic quantities and transport coefficients) in terms of
  the relativistic fluid variables. Here we mention the final
  expression for stress tensor, energy current and charge current:
\bea{
	t^{ij}	&= v^i v^j\rho + p g^{ij} -n\sigma^{ij} -z g^{ij} \nabla_k v^k,\nn \\
	j^i	&= v^i \left(\epsilon + p +\frac{1}{2}\rho
          \mathbf{v}^2\right)	-n \sigma^{ik} v_k -zv^i\nabla_k
        v^k\nn\\ 
		&\quad -\kappa\nabla^i \t - \t\s_I \nabla^i\left(\frac{\mu_I}{\t}\right)
		+ \s_I (\epsilon^i_I - v_j\beta^{ji}_I),\nn\\
	j^i_I &= q_I v^i 
	- \tilde\k_I\N^i \t 
	- \tilde\xi_{IJ} \N^i\lb\frac{\mu_J}{\t}\rb
	-\tilde m_I\N^i p \nn \\
	&+ \tilde\s_{IJ}\lb\epsilon^i_{J} - v_k \beta^{ki}_J \rb
	+ \Big\{\bar\k_I\e^{ij}\N_j \t
	+ \bar\xi_{IJ}\e^{ij} \N_j\lb\frac{\mu_J}{\t}\rb\nn \\
	& -\bar m_I\e^{ij} \N_j p 
	+ \bar\s_{IJ}\e^{ij}\lb\epsilon_{Jj} - v^k \beta_{Jkj} \rb \Big\}.
}
All the non-relativistic transport coefficients are fixed in terms of
relativistic transport coefficients and other thermodynamic
variables. Explicit relations have been provided in the main text.

\item
  In charge current we identify the thermal Hall contribution
  $\bar\k_I\e^{ij}\N_j \t$ (where $\tau$ is the temperature of 
  non-relativistic fluid). The Hall energy flow sourced by the
  temperature gradient is known as the Leduc-Righi effect. It has been
  predicted in condensed matter physics, that this kind of effect can
  be observed in various topological insulators. 

\item
  We also obtain the electromagnetic Hall energy current (parity-odd)
  $\bar\s_{IJ}\e^{ij}\lb\epsilon_{Jj} - v^k \beta_{Jkj} \rb$, where
  $\epsilon_{Jj}$ is the applied electric field and $\beta_{Iij}$ is
  proportional to the applied magnetic field (check \eq{E:elecfield}
  for exact definitions).

\item
We find that the non-relativistic fluid system satisfies the famous
`Wiedemann-Franz Law'.

\item
  Finally, we compute the entropy current $j_S^i$ for the
  non-relativistic fluid in \sctn{sec:non-rel-entrop-current}, and
  demand that the rate of entropy production of the system is always
  positive definite. For parity-even fluid it turns out to be quite
  straight forward. The canonical form of the entropy current confirms
  that the total entropy always increases due to dissipation, thermal
  and charge conduction. However for parity-odd fluid, canonical form
  of the entropy current fails to confirm the second law of
  thermodynamics. Therefore, we need to add two parity-odd vectors to
  the entropy current with arbitrary coefficients. Upon demanding the
  validity of second law, we see that one can fix these two
  coefficients exactly.

\item
Finally we observe that, unless the fluid is incompressible, all the
parity-odd transport coefficients must vanish to satisfy the second law of
thermodynamics. 

\end{itemize}


\section{Relativistic anomalous charged fluid}\label{sec:relflu}

Hydrodynamic description does not follow from the action principle,
rather it is generally formulated in the language of equations of
motion. The reason for this is the presence of dissipation in thermal
media. Due to internal friction called viscosity, a dissipative fluid
loses its energy over time as it propagates. The fluid without any
viscous drag is called an ideal fluid. In the simplest case, the
hydrodynamic equations (for a uncharged fluid) are the laws of
conservation of energy and momentum,
\be \label{E:conseqn1}
\grad_{\mu} T^{\mu\nu} =0 .
\ee
A relativistic fluid in $(d+1,1)$ dimensions ($d+1$ spatial and $1$
time dimensions) has total $d+3$ independent intensive fluid
variables: \emph{temperature} $(T)$, \emph{pressure} $(P)$ and
\emph{velocities} $u^\mu$ (only $d+1$ components of the velocity are
independent due to normalization $u^\mu u_\mu = -1$). Extensive
variables \emph{energy density} $(E)$ and \emph{entropy density}
$(S)$ are not considered to be independent as they can be determined
in terms of the intensive variables, by the first law of
thermodynamics and the equation of state.

Additionally we have an Euler's relation which follows from the
extensivity of the internal energy:
\be \label{E:thermoEOS}
	E+P = TS,
\ee
which in conjunction with the first law relates $P$ and $T$ as:
\be \label{E:thermoDiffPunQ} \df P = S\df T.  
\ee 
Therefore, among these thermodynamic variables, we can consider only
one to be independent, which we choose to be $T$. Thus, fluid
is determined in terms of $d+2$ variables: $u^\mu, T$. On the other
hand we have $d+2$ constitutive equations (\ref{E:conseqn1}), and
hence fluid system is completely determined. We can express the energy
momentum tensor $T^{\mu\nu}$ as a function of temperature, velocity
and their derivatives.

%
%
%

In case of fluid with multiple conserved $U(1)$ currents, we have more
fluid variables: \emph{charge densities} $(Q_I)$. The corresponding
conjugate variables are called \emph{chemical potentials} denoted by
$(M_I)$. In presence of conserved charges the first law is modified to
\be \label{E:thermoDiffE}
	\df E = T\df S + M_I\df Q_I.
\ee
From this equation one can compute the chemical potentials of the fluid
in terms of charges and other variables. However, in our computation
we consider chemical potentials $M_I$ to be our independent variables,
not the charges.
%
%
As a result, number of fluid variables have been increased to
$d+3$. However, in this case we also have another constitutive
equation (which follows from the global $U(1)$ symmetries),
\be
\N_{\mu} J^{\mu}_I =0.
\ee
Euler's relations are also modified to:
\be \label{E:thermoEOS}
	E+P = TS + M_I Q_I,
\ee
and from the first law one can write,
 \be \label{E:thermoDiffP} \df P
= S\df T + Q_I\df M_I.  
\ee
Therefore, for charged fluid we consider our basic thermodynamic
variables to be $T$ and $M_I/T$, rest are fixed in terms of these two
variables. We can therefore write the last equation in the following
form:
\be \label{E:PTMbyT} 
\df P = \frac{E+P}{T}\df T + TQ_I\df
\left(\frac{M_I}{T}\right).  
\ee
Thus, for a charged fluid, we express energy-momentum tensor and
charge current as a function of fluid velocities, temperature,
chemical potential and their derivatives. \\

One striking feature of a relativistic quantum field theory is
triangle anomalies. The effect of this anomaly is reflected in three
point correlation function of charge currents. However, the anomaly
does not affect the conservation of the current associated with a
global symmetry. When we put the theory in external background gauge
fields coupled to the currents, some of the currents will no longer be
conserved. \\

In the next subsection we discuss about the form of the energy
momentum tensor and global $U(1)$ current of a relativistic anomalous
fluid in presence of background gauge fields.


\subsection{Anomalous charged fluid: energy-momentum tensor and
  current}  

We review a generic charged relativistic fluid in $(d+2)$-dim with
anomalies. The \textit{energy-momentum tensor} is given by:
\be \label{E:TmunuDef}
	T^{\mu\nu}=(E+P)u^\mu u^\nu + P g^{\mu\nu} + \Pi^{\mu\nu},
\ee
where,
\be \label{E:PimunuDef}
	\Pi^{\mu\nu} = -2 \eta \tau^{\mu\nu} - \zeta \theta^{\mu\nu},
\ee
and fluid velocities are normalized 
\be \label{E:unorm}
	u^\mu u_\mu = -1.
\ee
Here, $\eta$ and $\zeta$ are relativistic shear and bulk viscosity
coefficients respectively. Up to first order in derivatives, the
respective terms are given by:
\be \label{E:taumunuDef}
	\tau^{\mu\nu}=\frac{1}{2}P^{\mu\alpha}P^{\nu\beta}\left[
          \nabla_\alpha u_\beta + \nabla_\beta u_\alpha -
          \frac{2}{d+1} g_{\alpha\beta}\theta \right], 
\ee
\be \label{E:thetaDef}
	\theta^{\mu\nu}=\theta P^{\mu\nu},\quad\quad \theta =
        \nabla_\alpha u^\alpha, 
\ee
where we use the projection operator:
\be \label{E:PmunuDef}
	P^{\mu\nu}=g^{\mu\nu}+u^\mu u^\nu.
\ee

Similarly \emph{charge current} of a relativistic fluid in presence of
multiple charges $Q_I$ ($I=1,2,3$) is given by\footnote{See
  \cite{Rangamani:2009xk} for a beautiful review.}
\be \label{E:JmuDef}
	J^{\mu}_{I}=Q_{I} u^\mu + \Upsilon^\mu_{I},
\ee
where,
\bem{ \label{E:UpsilonmuDef}
	\Upsilon^\mu_{I} =
	-\varrho_{IJ}P^{\mu\nu}\nabla_\nu \left(\frac{M_J}{T}\right) 
	+ \lambda_{IJ} E^\mu_J 
	- \gamma_I P^{\mu\nu}\nabla_\nu T \\
	+ \lbr \mho_I l^\mu 
	+ \tilde{\mho}_{IJ} B^\mu_J \rbr,
}
\be \label{E:lmuDef} l^\mu = \epsilon^{\mu \alpha \beta
  \gamma}u_\alpha\nabla_\beta u_\gamma.  
\ee
Here $\varrho_{IJ}$, $\lambda_{IJ}$ and $\g_I$ are charge, electric and thermal\footnote{Thermal
  conductivity of relativistic theory is fundamentally different from
  the thermal conductivity of non-relativistic theory. Entropy
  positivity demands relativistic thermal conductivity to vanish,
  however in non-relativistic limit it origins from
  elsewhere.} conductivities respectively. However, demanding the
positivity of local entropy current one can show show that, $\g_I$ vanishes and
$\varrho_{IJ}$ is related to $\lambda_{IJ}$.

We have kept the fluid in some background electromagnetic gauge fields
given by $A^\mu_I$:
\be \label{E:FieldIntro}
	F^{\mu\nu}_I = \nabla^\mu A^\nu_I - \nabla^\nu A^\mu_I,
\ee
\be \label{E:ElecField}
	E^\mu_I = F^{\mu\nu}_I u_\nu, \quad
	B^\mu_I = \frac{1}{2}\epsilon^{\mu\nu\alpha\beta}u_\nu F_{I\alpha\beta}.
\ee
The last two terms of \eq{E:UpsilonmuDef} are the most generic
single-derivative parity-odd modifications to the charge current
allowed in Landau frame. These terms are specific to
$(3+1)$-dimensions. In higher dimensions, the parity odd terms appear
only at higher derivative orders. However, we would like to keep our
calculations generic to $(d+2)$-dim, therefore in all expressions, we
represent parity-odd terms in curly braces $\lbr\dots\rbr$, with an
understanding that these terms contribute only for $d=2$.

We are working in Landau frame, which demands
\be
	u_\mu \Pi^{\mu\nu} = 0, \quad \quad u_\mu \Upsilon^\mu_I = 0.
\ee
and hence:
\be \label{E:TJnorm}
	u_\mu T^{\mu\nu} = -E u^\nu,  \quad\quad u_\mu J^\mu_I = - Q_I.
\ee
\ \\
The constitutive equations for this fluid are given by:
\be \label{E:TmunuCons}
	\nabla_\mu T^{\mu\nu}=F^{\nu\lambda}_IJ_{I\lambda},
\ee
\be \label{E:JmuCons}
	\nabla_\mu J^\mu_I = \lbr C_{IJK} E^\mu_J B_{K\mu} \rbr.
\ee
$C_{IJK}$ is called anomaly coefficient, which is completely symmetric in all indices. \\

Leading order $T^{\mu\nu}$ and $J^\mu$ must obey the first order
conservation equations, which can be found trivially using
Eqn. (\ref{E:TmunuCons}), (\ref{E:JmuCons}):
\bea{
  u^\alpha \nabla_\alpha E &= - (E+P)\theta, \label{E:FOE-E} \\
  P^{\mu\alpha}\nabla_{\alpha}P - Q_I E^\mu_I &= - (E+P) u^\alpha
  \nabla_\alpha u^\mu, \label{E:FOE-P} \\ 
  u^\mu\nabla_\mu Q_I + Q_I \theta &= \lbr C_{IJK} E^\mu_J B_{K\mu}
  \rbr. \label{E:FOE-Q} }
We have projected the equations along and perpendicular to the
direction of velocity for later convenience. We have also used
$E^\mu_I u_\mu=0$, which can be seen trivially. Using these,
$\tau^{\mu\nu}$ from Eqn. (\ref{E:taumunuDef}) can be written as:
\be 
\label{E:taumunuUseful} \tau^{\mu\nu} =\frac{1}{2}\left[
  \mathbf{Y}^{\mu\nu} + \mathbf{Y}^{\nu\mu} - \frac{2}{d+1}
  g^{\mu\nu}\theta - \mathbf{Z}u^\mu u^\nu \right], 
\ee
where,
\bea{ 
	\mathbf{Z} &= \frac{2}{d+1}\frac{u^\alpha\nabla_\alpha
          \mathbf{T}}{(E+P)}, \label{E:ZDef} \\ 
	\mathbf{T} &=(d+1)P-E, \label{E:TDef} \\
	\mathbf{Y}^{\mu\nu} &= \nabla^\mu u^\nu -
\frac{u^\nu \nabla^\mu P}{E+P} +Q_I\frac{u^\nu E^\mu_I}{E+P}.  \label{E:YmunuDef}
}
The trace of the energy-momentum tensor is given by:
\be \label{E:traceDef}
	T^{\mu}_{\ \mu}=\mathbf{T} -\zeta (d+1)\theta.
\ee
For conformal fluids $T^\mu_{\ \mu}=0$, which can be reached by
setting $\mathbf{T}=\zeta=0$. \\

Later, we shall use some relations involving $\mathbf{Y}^{\mu\nu}$,
which can be found trivially using Eqn. (\ref{E:YmunuDef}):
\be \label{E:u.Y}
	u_\mu\mathbf{Y}^{\mu\nu} = - \frac{ 2u^\nu u^\mu\nabla_\mu
          P}{E+P} - \frac{\nabla^\nu P}{E+P} + \frac{Q_I
          E^\nu_I}{E+P}, 
\ee
\be \label{E:Y.u}
	u_\nu\mathbf{Y}^{\mu\nu} = \frac{\nabla^\mu P}{E+P} -\frac{Q_I
          E^\mu_I}{E+P}.
\ee

\subsection{Entropy Current}

In this sub-section we write down the expression for entropy current
of a relativistic charged anomalous fluid and derive constraints on
various transport coefficients which follow from second law of
thermodynamics. The canonical form of relativistic entropy current is
given by
%
%
\be \label{E:JmuSDef}
	J^\mu_S	= S u^\mu - \frac{M_I}{T}\Upsilon^\mu_I, 
\ee
Using conservation equations Eqn. (\ref{E:TmunuCons}),
(\ref{E:JmuCons}) and thermodynamics Eqn. (\ref{E:thermoEOS}),
(\ref{E:thermoDiffP}) we can write:
\bem { \label{E:entropyRel1}
	\nabla_\mu J^\mu_S	
	= -\frac{1}{T}\Pi^{\mu\nu}\nabla_\mu u_\nu 
	+
        \left[\frac{E_{I\mu}}{T}-\nabla_\mu\left(\frac{M_I}{T}\right)
        \right]\Upsilon^\mu_{I}.
}
Using Eqn. (\ref{E:PimunuDef}), (\ref{E:UpsilonmuDef}) it can be
written (up to second order in derivative) as
\bem{ \label{E:entropyRel}
  \nabla_\mu J^\mu_S =
  \frac{1}{T}\eta\tau^{\mu\nu}\tau_{\mu\nu}+\frac{1}{T}\zeta \theta^2
  \\ 
  + \frac{1}{T}E_{I\mu}\lambda_{IJ} E^\mu_J +
  \left[P_{\alpha\mu}\nabla^\mu\left(\frac{M_I}{T}\right)\right]
  \varrho_{IJ}\left[P^{\alpha\nu}\nabla_\nu
    \left(\frac{M_J}{T}\right) \right] \\
  -\frac{1}{T}E^\mu_{I}\left(\varrho_{IJ}+T\lambda_{JI}\right)\nabla_\mu
  \left(\frac{M_J}{T}\right)  \\
  - \frac{1}{T}\gamma_I E^\mu_{I} \nabla_\mu T
  + \nabla_\mu\left(\frac{M_I}{T}\right)\gamma_I P^{\mu\nu}\nabla_\nu T \\
  + \bigg\{ - C_{KIJ} \frac{M_K}{T} E_{I\mu}B^\mu_J \\
	+ \left(\mho_I l^\mu + \tilde{\mho}_{IJ} B^\mu_J \right)
        \left[ \frac{1}{T} E_{I\mu} - \nabla_\mu\left(\frac{M_I}{T}\right) \right]\bigg\}.
}
Demanding entropy positivity, $\nabla_\mu J^\mu_S \geq 0$ we will get
\bea{ \label{E:Rconstraints1}
	\gamma_I &= 0, \quad
	\eta \geq 0, \quad \zeta \geq 0,\quad
	\lambda_{IJ}=\frac{1}{T}\varrho_{IJ}, \nn \\
	&\varrho_{IJ} \quad \text{matrix is positive definite}.
}
However the last two terms of \eq{E:entropyRel} cannot be made
positive definite. One would then expect that coefficients of those
terms must vanish! But in \cite{Son:2009tf}, the authors identified
that Eqn (\ref{E:entropyRel1}) is not the most generic expression for
the entropy current. They modified the entropy current with the most
generic parity odd vectors allowed by the symmetry
\be \label{E:JmuSMod}
	J^\mu_S	\ra J^\mu_S + \lbr Dl^\mu + \tilde{D}_IB^\mu_I \rbr.
\ee
In presence of these two terms the parity-odd part of the last
equation becomes,
\bem{
	\lbr \nabla_\mu \left(Dl^\mu + \tilde{D}_I B^\mu_I\right)
	- C_{KIJ} \frac{M_K}{T} E_{I\mu}B^\mu_J \dbrk
	+ \left(\mho_I l^\mu + \tilde{\mho}_{IJ} B^\mu_J \right)
        \left[ \frac{1}{T} E_{I\mu} -
          \nabla_\mu\left(\frac{M_I}{T}\right) \right] \rbr \\
	=
	\lbr l^\mu \left[ \nabla_\mu D
	- \frac{2D}{E+P}\nabla_\mu P
	- \mho_I\nabla_\mu\left(\frac{M_I}{T}\right)\right] \dbrk
	+ l^\mu E_{I\mu}\left[D\frac{2Q_I}{E+P} - 2\tilde{D}_I + \frac{1}{T} \mho_I\right]  \dbrk
	+ B^\mu_J \left[\nabla_\mu \tilde{D}_J
	- \frac{\tilde{D}_J}{E+P}\nabla_\mu P
	- \tilde{\mho}_{IJ}\nabla_\mu\left(\frac{M_I}{T}\right)\right]\dbrk
	+ E_I^\mu B_{J\mu}\left[\tilde{D}_J\frac{Q_I}{E+P} 
	+ \frac{1}{T}\tilde{\mho}_{IJ}
	- C_{KIJ} \frac{M_K}{T} \right] \rbr = 0.
}
Demanding the positivity of divergence of local entropy current we get,
%
\bea{
	\frac{\partial D}{\partial P} =
	\frac{2D}{E+P}, &\quad\quad
	\frac{\partial D}{\partial (M_I/T)} =\mho_I, \nn \\
	\frac{\partial \tilde{D}_J}{\partial P} = \frac{\tilde{D}_J}{E+P}, &\quad\quad
	\frac{\partial \tilde{D}_J}{\partial (M_I/T)} =\tilde{\mho}_{IJ}, \nn
}
\bea{\label{E:anomalyRelDE}
	&D\frac{2Q_I}{E+P} - 2\tilde{D}_I  = -\frac{1}{T}
	\mho_I, \nn \\
	&\tilde{D}_J\frac{Q_I}{E+P} = -\frac{1}{T}\tilde{\mho}_{IJ}+ C_{KIJ}
	\frac{M_K}{T}. 
}
One can in principle solve the above set of equation to get relations
between $D, \tilde D_I, \M_I, \tilde \M_{IJ}$ and $C_{IJK}$. \\

Finally we have the desired form of entropy current
\bem{ \label{E:entropyRel2}
	\nabla_\mu J^\mu_S =
        \frac{1}{T}\eta\tau^{\mu\nu}\tau_{\mu\nu}+\frac{1}{T}\zeta
        \theta^2 \\
+ \left[\frac{E^\alpha_{I}}{T}-P^{\alpha\mu}\nabla_\mu
          \left(\frac{M_I}{T}\right)\right] \varrho_{IJ}
        \left[\frac{E_{J\alpha}}{T}-P_{\alpha\nu}\nabla^\nu
          \left(\frac{M_J}{T}\right) \right],
}
which is entirely positive definite.


\section{Non-relativistic charged fluid} \label{sec:non-rel-fluid}

In this section, we review the properties of charged non-relativistic
dissipative fluid living in (d+1)-dimensions, in presence of some
background gauge fields. A non-relativistic fluid has following
constitutive equations:

\vspace{.5cm}

\noindent
\underline{{\it Continuity equation}}:
\be \label{E:continuity}
	\partial_t \rho + \partial_i (\rho v^i) = 0,
\ee
where $\rho$ is density of fluid particle.\\

\noindent
\underline{{\it Momentum conservation equation}}:
\be \label{E:momentumCons}
	\partial_t (\rho v^j) + \partial_i (t^{ij})=0,
\ee
where $t^{ij}$ is the energy-momentum tensor.\\

\noindent
\underline{{\it Energy conservation equation}}:
\be \label{E:energyCons}
	\partial_t \left(\epsilon+\frac{1}{2}\rho \mathbf{v}^2\right)
        + \partial_i j^{i}=0,
\ee
where $j^i$ is the energy current.\\

\noindent
\underline{{\it Conservation of charge current}}:
\be \label{E:chargeCons}
	\partial_t q_I + \partial_i j^i_I = 0,
\ee
where, $q_I$'s are the $U(1)$ charge densities and $\vec j_I$'s are the
corresponding currents. Note that energy current is denoted by $\vec j$
where as charge current is denoted by $\vec j_I$. \\

In $d$ spatial dimensions, charged fluid has total $d+3$ independent
variables (pressure $p$, temperature $\t$, chemical potentials $\mu_I$
and velocity $\vec v$), while others (density $\rho$, energy density
$\e$, charge density $q_I$ and entropy density $s$) are fixed in terms
of these using thermodynamic relations. Unlike the relativistic case
we do not consider the Euler's relation here. Later we shall see that
the light-cone reduction does not preserve extensivity of the fluid
system. Hence, the fluid is determined by $d+3$ parameters chosen to
be: $\t$, $\mu_I/\t$, $p$ and $\vec v$ (same as (d+1,1)-dim
relativistic fluid). On the other hand we have $d+3$ constitutive
equations.


Therefore, we can express energy-momentum tensor and charge
current in terms of $\t$, $\mu_I/\t$, $p$, $\vec v$ and
their derivatives. For ideal fluid they are given by,
\ben
	t^{ij} &=& \rho v^i v^j + pg^{ij}, \nn \\
	j^i &=& \left(\epsilon + p +\frac{1}{2}\rho \mathbf{v}^2\right)v^i, \nn\\
	j^i_I &=& q_I v^i. 
\een
\subsection{Non-relativistic dissipative fluid in background
  electromagnetic field}

When we put a dissipative fluid in background electromagnetic fields,
the above equations are modified. Continuity equation and charge
conservation equation remain the same but other two equations are
modified to
\bea{
  \partial_t (\rho v^j) + \partial_i(t^{ij})
  &=q_I\epsilon^j_I-j_{Ii}\beta^{ij}_I,\nn\\
	\partial_t \left(\epsilon+\frac{1}{2}\rho \mathbf{v}^2\right)
        + \partial_i j^{i} &=j^i_I\epsilon_{Ii},
}
where,
\be \label{E:elecfield}
	\epsilon^i_I = -\partial^i\phi_I -\partial_t a^i_I,\quad
	\beta^{ij}_I = \partial^i a^j_I-\partial^j a^i_I,
\ee
are electric and magnetic fields respectively. $\phi_I$ and $a^i_I$ are
scalar and vector potentials respectively.

For dissipative fluid the stress-energy tensor is given by,
\be \label{E:piijDef}
	t^{ij}=\rho v^i v^j + pg^{ij} + \pi^{ij},
\ee
where $\pi^{ij}$ is the correction to ideal fluid stress tensor due to
dissipation 
\be \pi^{ij} = - n\sigma^{ij}-z\delta^{ij}\partial_k v^k,
\ee 
\be \label{E:sigmaijDef} \sigma^{ij}=\partial^i v^j
+ \partial^j v^i - \delta^{ij}\frac{2}{d} \partial_k v^k. 
 \ee 
 $n$ and $z$ are the non-relativistic shear and bulk viscosity
 coefficients respectively.

The energy current for dissipative fluid is given by,
\be \label{E:jiDef}
	j^i=\left(\epsilon + p +\frac{1}{2}\rho \mathbf{v}^2\right)v^i + \vs^i,
\ee
where,
\ben\label{E:originalEcurrentcorrec}
	\vs^i &=&
	-n\sigma^{ij}v^j
	-z\partial_k v^kv^i
	-\kappa\partial^i \t \nn\\
&&	- \xi_I \nabla^i\left(\frac{\mu_I}{\t}\right)
	+ \sigma_I \epsilon^i_I
	- \a_I v_j\beta^{ji}_I.
\een
Here $\kappa$, $\xi_I$, $\s_I$ and $\a_I$ are thermal, charge,
electric and magnetic conductivities respectively. Later we shall see
that, for non-relativistic fluid, obtained by light cone reduction of a
relativistic fluid, $\xi_I$ and $\a_I$ are
related to $\s_I$. \\

Similarly charge current for a dissipative fluid is given by,
\be
j_I^i = q_I v^i + \vs_I^i,
\ee 
where,
\bem{\label{E:originalQcurrentcorrec}
	\vs_I^i = 
	- \tilde\k_I\N^i \t 
	- \tilde\xi_{IJ} \N^i\lb\frac{\mu_J}{\t}\rb
	-\tilde m_I\N^i p
	+ \tilde\s_{IJ}\e^i_{J} \\
	- \tilde\a_{IJ}v_k\beta^{ki}_{J}
	+ \Big\{\bar\k_I\e^{ij}\N_j \t
	+ \bar\xi_{IJ}\e^{ij} \N_j\lb\frac{\mu_J}{\t}\rb\\
	-\bar m_I\e^{ij} \N_j p 
	+ \bar\s_{IJ}\e^{ij}\e_{Jj}
	- \bar\a_{IJ}\e^{ij}v^k\beta_{Jkj} \Big\}.
}
$\tilde\k_I,\tilde\s_{IJ},\tilde\xi_{IJ},\tilde m_I,\tilde \a_{IJ}$
and $\bar\k_I,\bar\s_{IJ},\bar\xi_{IJ},\bar m_I,\bar \a_{IJ}$ are some
arbitrary parity-even and parity-odd transport
coefficients\footnote{The sign of these coefficients are completely
  arbitrary for now, and are chosen keeping in mind later
  convenience.}. Again, for a non-relativistic fluid, obtained by
light-cone reduction these transport coefficients are fixed in terms
of relativistic transport coefficients. Also, they are constrained
when we demand positivity of local entropy current.

\section{Light-cone reduction} \label{sec:lcr}

Discrete light cone quantization of a $(d+1,1)$-dim
relativistic quantum field theory  (QFT) boils down to a non-relativistic
quantum field theory in one lower dimension. Let us consider a QFT in flat
spacetime with metric,
\be
ds^2 = -(dx^0)^2 +(dx^{d+1})^2 + \sum_{i=1}^{d} (dx^i)^2 .
\ee
We introduce the light-cone coordinates,
\be
	x^{\pm} = \frac{1}{\sqrt2}\lb x^0 \pm x^{d+1} \rb.
\ee
In the light-cone frame the metric can be written as,
\be
ds^2 = - 2 dx^+dx^-  + \sum_{i=1}^{d} (dx^i)^2 .
\ee
Suppose we view the QFT in this light-cone frame and evolve it in
light-cone time $x^+$. Then, for fixed light-cone momentum $P_-$, we
obtain a system in $d + 1$ dimensions with non-relativistic
invariance. This is because the symmetry algebra of the relativistic
theory reduces to corresponding non-relativistic symmetry algebra upon
light-cone reduction. For example, if the relativistic theory is
invariant under Poincar\'e transformation, then the corresponding
algebra ($SO(d+1,1)$) reduces to $d$ dimensional Galilean algebra. On
the other hand if we consider a QFT with conformal invariance in
$(d+1)+1$ dimensions, then upon light cone reduction the corresponding
algebra ($SO(d+2,2)$) reduces to Schr\"odinger algebra in $d$ spatial
dimensions. This is known as discrete light-cone quantization of
quantum field theories.

Since hydrodynamics is low energy fluctuation of equilibrium quantum
field theory, light-cone reduction of relativistic constitutive
equations boil down to the non-relativistic constitutive equations for
a fluid in one lower dimension. Relativistic charged fluid in
$(d+1,1)$-dim, as we have already discussed, has $d+3$ degrees
of freedom: temperature, chemical potential and normalized
velocities. On the other had, a non-relativistic fluid in $d$ spatial
dimensions also has total $d+3$ degrees of freedom: temperature,
pressure, chemical potential and velocities. Our goal is to consider
the most generic relativistic anomalous fluid system in presence of
background electromagnetic fields in $(3+1)$ dimensions and reduce the
constitutive equation (also equation for entropy current) over
light-cone coordinates and obtain the corresponding non-relativistic
equations (most generic) for a fluid in one lower dimensions. We also
find a mapping between the degrees of freedom of the
$(d+2)$-dimensional fluid to the degrees of freedom of the $(d +
1)$-dimensional fluid.

We denote the $d$ spatial coordinates with $x^i$. Metric components in
light-cone coordinates are given by
\be \label{E:LCRgij}
	g^{ij}=\delta^{ij} \quad\text{and}\quad g^{+-}=-1,
\ee
rest are zero. Gradient operator is given by,
\be \label{E:partialsDef}
	\nabla_{\mu}=\{ \partial_+,\partial_-,\partial_{i} \}
        \quad\text{and}\quad \nabla^{\mu}=\{
        -\partial_-,-\partial_+,\partial_{i} \}. 
\ee
As we have seen that not all components of $u^\mu$, $J^\mu_I$ and
$T^{\mu\nu}$ are independent, from Eq. (\ref{E:TJnorm}) we get,
\bea{ 
	u^- 		&= \frac{1}{2u^+}(1+\mathbf{u}^2), \label{E:u-Def} \\ 
	u^+ J^-_I  	&= Q_I + u_i J^i_I - u^- J^+_I, \label{E:J-Def} \\
	u^+ T^{\mu -} 	&= E u^\mu + u_i T^{\mu i} - u^- T^{\mu
          +}. \label{E:Tmu-Def} 
}
We shall reduce the theory along the $x^-$ direction, and consider
$x^+$ to be the non-relativistic time. We will consider only solutions
to the relativistic equations that do not depend on $x^-$; that is,
all derivatives $\pa_-$ vanish.

\subsection{Reduction of background fields}\label{rednofback}

We reduce the Maxwell's equations for the background fields of
relativistic fluid to get consistent background for non-relativistic
theory. Maxwell's equations for relativistic system are given by (we
shall assume that the sources for these background fields are far away
from the fluid),
\be
	\nabla_\mu F^{\mu\nu}_I = 0,
\ee
%

Under light-cone reduction the above equations take the following form:
\bea{
  \vec{\nabla}^2 A^+_I &= 0, \nn \\
  \nabla_i \left( \nabla^i A^-_I + \nabla_+ A^i_I \right) &=
  -\nabla^2_+ A^+_I, \nn \\
  \nabla_i \left(\nabla^i A^j_I - \nabla^j A^i_I \right) &=
  \nabla^j\nabla_+ A^+_I.  }
These equations can be identified with source free static Maxwell's
equations of a non-relativistic system if we map\footnote{In
\apndx{nrelectro} we have discussed how to expand Maxwell's equations
  in powers of $1/c$ to get the non-relativistic limit.}:
\ben\label{E:PotenIden}
	A^-_I &=& \phi_I \ \ (\text{scalar potential}),\nn \\
	A^i_I &=& a^i_I \ \ (\text{vector potential}),
\een
and
\bea{
	\nabla^i\nabla_i A^+_I = 0 , \ \ 
	\nabla_+\nabla_+ A^+_I = 0,  \ \
	\nabla^i\nabla_+ A^+_I = 0.
}
Which would inturn tell us
\bea{
	\nabla_+ A^+_I, \nabla_i A^+_I = \text{constant}.
}
We would however like $A^+_I$ to have some finite value at infinity,
which will enforce:
\bee{\label{E:A+const}
	A^+_I = \text{constant}.
}
With this identification, from Eq. (\ref{E:ElecField}) one can show that
\be \label{E:ElecIden}
	E^+_I = 0, \quad\quad
	E^-_I = u_i\epsilon^i_I, \quad\quad
	E^i_I = u^+ \epsilon^i_I - u_j\beta^{ji}_I,
\ee
and
\ben \label{E:MagIden}
	B^+_I = -u^+\epsilon_{ij}\beta^{ij}_I, &&\quad\quad
	B^-_I = u^-\epsilon_{ij}\beta^{ij}_I - 2\epsilon_{ij}u^i\epsilon^j_I, \nn\\
	B^i_I &=& -2u^+\epsilon_{ij}\epsilon^j_I,
\een
where, $\epsilon^i_I = -\pa^i\phi_I-\pa_ta_I^i$ and $\beta_I^{ij}=
\pa^ia_I^j-\pa^ja_I^i$. 

Finally let us check what happens to the gauge freedom in relativistic
side. The gauge transformation is give by,
\bee{ A^{\mu} \ra A^{\mu} + \N^\mu \L.  }
In the light-cone coordinates it becomes 
\bea{
	A^{+} &\ra A^{+}, \nn \\
	A^{-} &\ra A^{-} - \N_+ \L, \nn \\
	A^{i} &\ra A^{i} + \N^i \L.
}
Thus we see that the gauge freedom of relativistic theory reduces to
the gauge freedom of the non-relativistic theory. Additionally $A^+$,
which we have fixed to be a constant, under a gauge choice does not
change. Otherwise, one could perform a local gauge transformation to
modify $A^+$ to a non constant value, which would then break our
identification.

\subsection{Reduction of energy-momentum tensor and charge current}

The reduction of relativistic equations of energy-momentum and charge
conservation after using \eq{E:A+const} are given by:
\begin{align}
	\nabla_+ T^{++} + \nabla_i T^{i+} &= 0, \label{E:Tmu+ConsLCR} \\
	\nabla_+ T^{+-} + \nabla_i T^{i-} &= -J^i_I\left(\nabla_i
          A^-_I + \nabla_+ A_{Ii} \right), \label{E:Tmu-ConsLCR} \\ 
	\nabla_+ T^{+j} + \nabla_i T^{ij} &= -J^+_I\left(\nabla_+
          A^j_I + \nabla^j A^-_{I} \right) \nn \\
          &\quad -J^i_I\left(\nabla_i A^j_I
          - \nabla^j A_{Ii} \right), \label{E:TmujConsLCR} \\ 
	\nabla_+ J^+_I + \nabla_i J^i_I &= \lbr C_{IJK} E^\mu_J
        B_{K\mu} \rbr = 0. \label{E:JmuConsLCR} 
      \end{align}
      R.H.S. of Eqn. (\ref{E:JmuConsLCR}) can be shown to vanish
      explicitly after reduction in four dimensions.

These equations reduce to non-relativistic equations under following
identifications
\bea{
	T^{++}	= \rho, \qquad & T^{i+}=\rho v^i, \nn \\
	T^{+-}	= \epsilon + \frac{1}{2}\rho \mathbf{v}^2, \qquad
        T^{i-} &=j^i, \qquad T^{ij} = t^{ij}, \nn \\ 
	J^{+}_I	= q_I, \qquad & J^{i}_I=j^i_I.
} \label{Eq:identification}
The identification also tells that the non-relativistic charge current
is conserved, while the relativistic charge current was not. We shall
now attempt to use this mapping to establish relations between
relativistic and non-relativistic parameters.

\subsubsection{$T^{++}$ calculation}

\noindent
Using Eqn. (\ref{E:taumunuUseful}) and (\ref{E:thetaDef}) we can show:
\bea{
	\tau^{++}	&=- \frac{1}{2}(u^+)^2\mathbf{Z}, \nn \\
	\theta^{++}	&= (u^{+})^2\theta,
}
and thus:
\be \label{E:T++Calc}
	T^{++}	=(E+P)(u^+)^2
		+(u^+)^2\left(\eta\mathbf{Z}
		-\zeta \theta\right).
\ee
Identifying $T^{++}=\rho$
\be \label{E:rhoIdenrel}
	\boxed{\rho	=(E+P)(u^+)^2
			+(u^+)^2\left(\eta\mathbf{Z}
			-\zeta \theta\right)}.
\ee

\subsubsection{$T^{i+}$ calculation}

\noindent
Using Eqn. (\ref{E:taumunuUseful}) and (\ref{E:thetaDef}) we can show:
\bea{
	\tau^{i+}	&=\frac{1}{2}\mathbf{Y}^i  - \frac{1}{2}\mathbf{Z}u^i u^+, \nn \\
	\theta^{i+}	&=u^{i}u^{+}\theta,
}
where,
\be \label{E:YmuDef}
	\mathbf{Y}^\mu=\mathbf{Y}^{\mu+}=\left( \nabla^\mu u^+ -
          \frac{u^+\nabla^\mu P}{E+P} + u^+\frac{Q_IE^\mu_I}{E+P}
        \right).
\ee
Using Eqn. (\ref{E:rhoIdenrel}), we will find,
\be \label{E:Ti+Calc}
	T^{i+}	= \frac{u^i}{u^+} \rho - \eta\mathbf{Y}^i.
\ee
Identifying $T^{i+}=\rho v^i$
\be \label{E:viDef}
	\boxed{v^i= \frac{u^i}{u^+} -\frac{\eta}{\rho}\mathbf{Y}^i}.
\ee

\subsubsection{$T^{ij}$ calculation}

\noindent
Using Eqn. (\ref{E:taumunuUseful}) we will find:
\begin{align}
	\tau^{ij}	&=\frac{1}{2}\left[ \mathbf{Y}^{ij} +
          \mathbf{Y}^{ji} - \frac{2}{d+1} g^{ij}\theta - \mathbf{Z}u^i
          u^j \right]. 
\end{align}
It's easy to check that:
\be \label{E:thetadvkrelation}
	-\frac{u^\alpha \nabla_\alpha E}{(E+P)}	= \theta =
        -\frac{u^\alpha \nabla_\alpha P}{(E+P)}+u^+ \nabla_k v^k, 
\ee
and thus introducing $\mathbf{Z}$ from Eqn. (\ref{E:ZDef}),
\be
	-\frac{2}{(d+1)}\theta	= \frac{1}{d}\mathbf{Z}
        -u^+\frac{2}{d} \nabla_k v^k. 
\ee
Using the above relations and the identity:
\be
	\mathbf{Y}^{\mu i} = u^+ \nabla^\mu v^i + v^i \mathbf{Y}^\mu,
\ee
we find that:
\be
	\tau^{ij}	=\frac{1}{2}\left(u^+\sigma^{ij} + v^i\mathbf{Y}^j + v^j\mathbf{Y}^i
				+ \frac{g^{ij}}{d}\mathbf{Z}
				- (u^+)^2 v^i v^j\mathbf{Z}\right),
\ee
where $\sigma^{ij}$ is given by Eqn. (\ref{E:sigmaijDef}). Similarly
using Eqn. (\ref{E:thetaDef}) we can reach to:
\be
	\theta^{ij}	=-g^{ij}\frac{u^\alpha \nabla_\alpha P}{(E+P)}
			+g^{ij}u^+\nabla_{k}v^k
			+(u^+)^2v^{i}v^{j}\theta,
\ee
and thus:
\ben \label{E:TijCalc}
	T^{ij} &=& v^i v^j\rho
		+ g^{ij} \left[ P 
		-\left(
			\frac{\eta}{d}\mathbf{Z}
			-\zeta \frac{u^\alpha \nabla_\alpha P}{E+P}
		\right)
		\right]\nn\\
		&& \quad -\eta u^+\sigma^{ij}
		-\zeta u^+ g^{ij} \nabla_k v^k.
\een
Identifying $T^{ij}=t^{ij}$
\be \label{E:pDef}
	\boxed{p = P -\left(\frac{\eta}{d}\mathbf{Z}-\zeta
            \frac{u^\alpha \nabla_\alpha P}{E+P}	\right)},
\ee
\be \label{E:nDef}
	\boxed{n=\eta u^+},
\ee
\be \label{E:zDef} \boxed{z=\zeta u^+}.
\ee

\subsubsection{$T^{+-}$ calculation}

\noindent
Using Eqn. (\ref{E:Tmu-Def}), (\ref{E:T++Calc}) and (\ref{E:Ti+Calc})
we can find:
\be \label{E:T+-Calc}
	T^{+-}	=\frac{1}{2}(E-P) 
		-\frac{1}{2}\left(\eta \mathbf{Z} - \zeta \theta \right)
		+ \frac{1}{2}\rho\mathbf{v}^2.
\ee
Identifying $\displaystyle T^{+-}=\epsilon+\frac{1}{2}\rho \mathbf{v}^2$
\be \label{E:eDef}
	\boxed{\epsilon=\frac{1}{2}(E-P) - \frac{1}{2}\left(\eta
            \mathbf{Z} - \zeta \theta \right)}. 
\ee

\subsubsection{$T^{i-}$ calculation}

\noindent
Using Eqn. (\ref{E:Tmu-Def}), (\ref{E:Ti+Calc}) and (\ref{E:TijCalc})
we can find: 
\be \label{E:Ti-Calc}
	T^{i-}	= v^i \left(\epsilon + p +\frac{1}{2}\rho \mathbf{v}^2\right)
		-n \sigma^{ik} v_k
		-zv^i\nabla_k v^k
		+\eta \frac{1}{(u^+)^2}\mathbf{Y}^i.
\ee
The last term can be written as:
\bem{
\label{E:YiThermo} \eta \frac{1}{(u^+)^2}\mathbf{Y}^i =
-\eta\frac{1}{T}\nabla^i \left(\frac{T}{u^+}\right) - \eta u^+
\frac{Q_I T}{\rho} \nabla^i\left(\frac{M_I}{T}\right) \\
+ \eta u^+\frac{Q_I}{\rho} E^i_I. 
}
We identify (leading order):
\be \label{E:tIden}
	\t =\frac{T}{u^+}, \qquad \mu_I = \frac{M_I}{u^+}, \qquad q_I =
        Q_I u^+ ,
\ee
and
\be \label{E:kappaIden}
	\kappa = \frac{\eta}{\t u^+} = 2n\frac{\epsilon + p}{\t
            \rho},\qquad \s_I = n \frac{q_I}{\rho}.
\ee
With this identification we get,
\be \label{E:YiThermoNR}
	\eta \frac{1}{(u^+)^2}\mathbf{Y}^i = -\kappa\nabla^i \t -
        \t\s_I \nabla^i\left(\frac{\mu_I}{\t}\right)
        + \s_I (\epsilon^i_I - v_j\beta^{ji}_I). 
\ee
Identifying $T^{i-}=j^i$ we have:
\ben
	j^i	&=& v^i \left(\epsilon + p +\frac{1}{2}\rho \mathbf{v}^2\right)
		-n \sigma^{ik} v_k
		-zv^i\nabla_k v^k\nn\\
		&& -\kappa\nabla^i \t - \t\s_I \nabla^i\left(\frac{\mu_I}{\t}\right)
		+ \s_I (\epsilon^i_I - v_j\beta^{ji}_I).\nn\\
\een
Comparing it with \reff{E:originalEcurrentcorrec}, we find that we
have the expected form. Also we have estabilished relations between
charge, electric and magnetic conductivities: \bee{ \frac{\xi_I}{\t} =
  \s_I = \a_I.  }

\subsubsection{$J^+$ calculation}

\noindent
Using Eqn. (\ref{E:UpsilonmuDef}) we can find
\ben
	\Upsilon^+_{I}	&=& -\varrho_{IJ}u^+u^\nu\nabla_\nu
        \left(\frac{M_J}{T}\right) - \gamma_I u^+u^\nu\nabla_\nu T
        \nn \\
       && +  \lbr \mho_I l^+ + \tilde{\mho}_{IJ} B^+_J \rbr, 
\een
\be
	l^+	= -\epsilon^{+-ij}(u^+)^2\nabla_i v_j =
        -\epsilon^{ij}(u^+)^2\nabla_i v_j. 
\ee
Using Eqn. (\ref{E:MagIden}) we can write
\ben \label{E:Upsilon+Def}
	\Upsilon^+ &=& - u^+\bigg[\varrho_{IJ}u^\nu\nabla_\nu
          \left(\frac{M_J}{T}\right) + \gamma_I u^\nu\nabla_\nu
          T\nn\\
 && +          \lbr\mho_I \epsilon^{ij}u^+\nabla_i v_j+
          \tilde{\mho}_{IJ}\epsilon_{ij}\beta^{ij}_J\rbr\bigg]. 
\een
Identifying $J^+_I = q_I$,
\ben \label{E:qIdenrel}
	\boxed{q_I=u^+Q_{I} + u^+ \vp_I},
\een
where,
\bem{ \label{E:qIdenrel}
	\vp_I=-\bigg[\varrho_{IJ}u^\nu\nabla_\nu
            \left(\frac{M_J}{T}\right) + \gamma_I u^\nu\nabla_\nu T \\
            + \lbr\mho_I \epsilon^{ij}u^+\nabla_i v_j+
            \tilde{\mho}_{IJ}\epsilon_{ij}\beta^{ij}_J\rbr\bigg]. 
}
This tells us the sub-leading correction to non-relativistic charges.

\subsubsection{$J^i$ calculation}

\noindent
Using Eqn. (\ref{E:UpsilonmuDef}) we can find
\bem{
	\Upsilon^i_{I}	
	=
	-\varrho_{IJ}P^{i\nu}\nabla_\nu \left(\frac{M_J}{T}\right) 
	+ \lambda_{IJ} E^i_J 
	- \gamma_I P^{i\nu}\nabla_\nu T \\
	+ \lbr \mho_I l^i 
	+ \tilde{\mho}_{IJ} B^i_J \rbr.
}
One can find using \eq{E:YiThermo} that:
\bee{
	\lbr l^i \rbr 	= \lbr v^i l^+ + (u^+)^2 \Xi^i\rbr, \label{E:l+liRel}
}
where
\bem{
	\lbr \Xi^i \rbr	=
	\lbr\epsilon^{ij}\left( \frac{1}{\t
            (u^+)^2}\nabla_j \t + \frac{q_I t}{\rho}
          \nabla_j\left(\frac{\mu_I}{\t}\right) -
          \frac{2}{\rho}\nabla_j p \tbrk
          + \frac{1}{\rho}Q_J E_{Jj}
        \right)\rbr.
}
Certain velocity dependent terms appear in the above expression, which
can be shown to sum to zero in 2 spatial dimensions. \\

Using thermodynamic relations Eqn. (\ref{E:thermoEOS}),
(\ref{E:thermoDiffP}), (\ref{E:YiThermo}), along with all light cone
identifications, we will find
\bem{ J^i_{I} = v^i J^+_I - \frac{\gamma_I u^+\t}{2(\epsilon +
    p)}\nabla^i p
  - \frac{\k q_I}{2(\e + p)}\nabla^i \t \\
  - \bigg(\varrho_{IJ} + \xi_I\frac{q_J}{2(\epsilon + p)} - \gamma_I
  u^+ \frac{\t^2 q_J}{2(\epsilon + p)} \bigg) \nabla^i
  \left(\frac{\mu_J}{\t}\right)\\
  + \bigg(\frac{\s_Jq_I}{2(\epsilon + p)} + \lambda_{IJ}u^+ \bigg)
  \left(\epsilon^j_{J} - v_k \beta^{kj}_{J} \right) \\
  - \lbr\mho_I (u^+)^2\frac{2}{\rho}\epsilon^{ij}\rbr\nabla_j p
  + \lbr\frac{\mho_I}{\t} \epsilon^{ij}\rbr\nabla_j \t\\
  + \lbr\mho_I (u^+)^2 \frac{q_J \t}{\rho} \epsilon^{ij}\rbr \nabla_j
  \left(\frac{\mu_J}{\t}\right) \\
  + \lbr\mho_I (u^+)^2 \frac{q_J}{\rho} \epsilon^{ij} -
  \tilde{\mho}_{IJ} u^+ 2\epsilon^{ij}\rbr \left(\epsilon_{Jj} - v^k
    \beta_{Jkj} \right).  }
We identify
\ben \label{E:rijIden} \tilde m_I &=& \frac{\gamma_I \t
  u^+}{2(\epsilon + p)}, \quad \tilde\xi_{IJ} = \left[\varrho_{IJ} +
  \frac{\xi_Jq_I}{2(\epsilon +
    p)} - m_Iq_J \t\right], \nn \\
\tilde\k_I &=& \frac{\k q_I}{2(\e + p)}, \quad \tilde\s_{IJ} =
\tilde\a_{IJ} = \left[\lambda_{IJ}u^+ +
  \frac{\s_Jq_I}{2(\epsilon + p)}\right], \nn \\
\bar m_I &=& \frac{2\omega_I}{\rho}, \qquad
\bar\xi_{IJ} = \xi_J\frac{\omega_I}{n}, \nn \\
\bar\k_I &=& \k\frac{\omega_I}{n},\qquad \bar\s_{IJ} = \bar\a_{IJ} =
\lB \s_J\frac{\omega_I}{n} - 2\tilde{\omega}_{IJ}\rB, \een where,
\ben \label{E:lijIden} \omega_I &=& \mho_I (u^+)^2, \qquad
\tilde{\omega}_{IJ} = \tilde{\mho}_{IJ} u^+.  \een
Finally we have the desired form of charge current
\be
	j^i_I = q_I v^i + \vs^i_I,
\ee
\bem{\label{E:upsilonDef}
	\vs_I^i = 
	- \tilde\k_I\N^i \t 
	- \tilde\xi_{IJ} \N^i\lb\frac{\mu_J}{\t}\rb
	-\tilde m_I\N^i p \\
	+ \tilde\s_{IJ}\lb\epsilon^i_{J} - v_k \beta^{ki}_J \rb
	+ \Big\{\bar\k_I\e^{ij}\N_j \t
	+ \bar\xi_{IJ}\e^{ij} \N_j\lb\frac{\mu_J}{\t}\rb\\
	-\bar m_I\e^{ij} \N_j p 
	+ \bar\s_{IJ}\e^{ij}\lb\epsilon_{Jj} - v^k \beta_{Jkj} \rb \Big\},
}
Note that all transport coefficients are not independent. The only
independent coefficients are: $\tilde m_I$, $\tilde \xi_{IJ}$,
$\tilde\s_{IJ}$, $\omega_I$ and
$\tilde{\omega}_{IJ}$. \\

\noindent \textbf{Wiedemann-Franz Law:} This famous law predicts that
the ratio of charge conductivity (which appears in charge
current)\footnote{Lets consider only one $U(1)$ charge here.} to
thermal conductivity (which appears in energy current) in metals as:
$\tilde\s/\k = 1/L\t$, where $L$ is the Lorenz number predicted to be
$\sim 2.45 \times 10^{-8} \ W \O K^{-2}$. The law is found to be in
good agreement with experiments. We attempt to check the same in our
setup\footnote{We have used here the fact that non-relativistic system
  respects the constraint $\t \tilde\s_{IJ} = \tilde\xi_{IJ}$, which
  has been showed in next section.}:
\bee{
	\frac{\tilde\s}{\k} = \frac{\rho\varrho}{2n(\epsilon + p)} +
        \frac{\t q^2}{4(\epsilon + p)^2}, 
}
We model the electrons in metals as free classical gas with no
external pressure: fluid with homogeneous particles each of charge $e$
(electronic charge), mass $m_e$ and average energy $3/2 k_B \t$. One
can check that under mentioned assumptions, our system follows
Wiedemann-Franz Law with Lorenz number given as:
\bee{
	L = \lb\frac{\varrho}{n}\frac{ m_e}{3 k_B} + \frac{e^2}{9k_B^2}\rb^{-1},
}
Assuming $\varrho$ and $n$ of nearly same order, first term turns out
to be about 15 orders of magnitude smaller than the second term and
can be safely neglected. Hence Lorenz number is our case is given
approximately by: $6.68 \times 10^{-8} \ W \O K^{-2}$, given our
assumptions, which is in fair agreement with the experimental value.

\subsection{Thermodynamics of the reduced system}

We start with a relativistic system with variables satisfying the
Eq. (\ref{E:thermoDiffE}) and (\ref{E:thermoEOS}). The first equation
is the first law of thermodynamics and the second equation is called
the Euler's equation. The Euler equation follows from the additive
property of internal energy which is a homogeneous function of degree
one. After light-cone reduction when we map non-relativistic variables
in terms of relativistic ones, we see that at the leading order (ideal
fluid), non-relativistic variables satisfy the following two
equations,
\ben\label{Eq:non-relthermo}
	\df \e = \tau \df s + \mu_I \df q_I &+& (\e+p -s\t - q_I
        \mu_I)\frac{\df\rho }{\rho}, \nonumber \\
2(\epsilon + p) &=& s \tau + \mu_I q_I .
\een
Here we have identified the non-relativistic entropy density $s=S
u^+$. The first equation, which follows from
Eq. (\ref{E:thermoDiffE}), is the first law of thermodynamics
satisfied by the non-relativistic variables. The second, coming from
Eq. (\ref{E:thermoEOS}), is not exactly the Euler's equation for
non-relativistic variables. We comment on this equation at the end of
this sub-section.

As we consider the first derivative corrections to non-relativistic
variables (evaluated in the previous sub-section), the first law of
thermodynamics changes. However, that can not be the case. If we
demand that our non-relativistic system is a physical one, then the
first law should be satisfied by the non-relativistic variables at
every order in derivative expansion. This demand forces us to add some
higher derivative corrections to entropy density which we could not
determine in \sctn{Eq:identification}. Let's assume the following
corrected forms:
\bea{ s = Su^+ (1+\chi), 
}
$\chi$ is corresponding corrections.

Demanding first law of thermodynamics to be satisfied at first order
in derivative we get:
\bem{\label{E:thermocorrec2} TS \df \chi +
  M_I \df \varpi_I
  - M_I \varpi_I\frac{\df E}{E+P}\\
  - \chi M_I \left(\df Q_I -
    \frac{Q_I}{E+P}\df E\right)
  =\\
  -\eta \frac{1}{d}\mathbf{Z}\frac{\df \mathbf{T}}{E+P} - \zeta\frac
  {\df E}{E+P} \frac{u^\alpha\nabla_\alpha E}{E+P}
  + \zeta\frac {\df (E-P)}{E+P} \frac{u^\alpha\nabla_\alpha (E-P)}{E+P} \\
  + \bigg[ \chi ST  + M_I \varpi_I + \eta
  \frac{d+2}{d}\mathbf{Z} \\
  - \zeta \left(\theta + 2 \frac{u^\alpha\nabla_\alpha P}{E+P}\right)
  \bigg] \lb \frac{\df P}{E+P} + \frac{\df u^+}{u^+} \rb. }
%
%
%
Note that, thermodynamic relations (\ref{E:thermoEOS}) -
(\ref{E:thermoDiffP}), conservation equations (\ref{E:FOE-E}) -
(\ref{E:FOE-Q}) up to first order and all the identifications between
non-relativistic and relativistic variables (obtained by LCR) have
been extensively used to carry out the above computation. Now, using
(\ref{E:FOE-E}) - (\ref{E:FOE-Q}), we can write the above equation in
the following form\footnote{There will be two terms proportional to
  $\lbr C_{IJK}E^\mu_J B_{K\mu} \rbr$ which we have checked will
  vanish after reduction.  }
\bem{\label{E:thermocorrec3}
	TS u^\mu\nabla_\mu \chi
	+ M_I u^\mu\nabla_\mu \varpi_I
	+ M_I \varpi_I\theta \\
	=
	-\eta \frac{d+1}{2d}\mathbf{Z}^2
	- \zeta\theta^2
	+ \zeta(u^+)^2\left(\nabla_k v^k\right)^2.
}
This equation fixes the correction to the entropy density.

The last equation in (\ref{Eq:non-relthermo}), which follows from
Euler's equation (for the relativistic variables) does not imply that
the non-relativistic energy density $\epsilon$ is a homogeneous
function of degree one. In fact, this equation gets modified with the
corrected values of non-relativistic entropy density\footnote{Although
  the form of this equation can be maintained by adding some one
  derivative correction to temperature or chemical
  potential.}. Therefore, we conclude that the non-relativistic system
we obtain after light-cone reduction of a homogeneous relativistic
fluid system, is not homogeneous any more (even at the leading order)
$i.e.$ the internal energy is not a homogeneous function of degree
one.


\section{Non-relativistic Entropy
  Current}\label{sec:non-rel-entrop-current} 


The goal of this section is to write down an entropy current for a
non-relativistic fluid system, and find constraints on various
transport coefficients from the positivity of local entropy
production. Using
\eq{E:thermocorrec3} and properties of $\mathbf{Y}^{\mu\nu}$ , one can
reduce the \eq{E:entropyRel} and after
a lengthy algebra, we  get
the following equation,
\begin{multline} \label{E:entropyNRfull}
	\nabla_+ s
	+ \nabla_i j^i_S
	+ \nabla_i \left(\frac{\eta}{\t (u^+)^2}\mathbf{Y}^i\right)
	=
	\frac{1}{\t}\frac{n}{2}\sigma^{ij}\sigma_{ij}
	+ \frac{1}{\t}z(\nabla_k v^k)^2 \\
	+ \kappa\left[\frac{1}{\t}\nabla^i \t + \frac{q_I
            \t}{2(\epsilon+p)} \nabla^i\left(\frac{\mu_I}{\t}\right) -
          \frac{q_I
            (\epsilon^i_{I}-v_k\beta^{ki}_{I})}{2(\epsilon+p)}\right]^2
        \\ 
	-
        \left[\frac{\epsilon^i_{I}-v_k\beta^{ki}_{I}}{\t}-\nabla^i\left(\frac{\mu_I}{\t}\right)\right] 
        \frac{\xi_Jq_I}{2(\epsilon + 
          p)}\left[\frac{\epsilon_{Ji}-v^k\beta_{Jki}}{\t}-\nabla_i\left(\frac{\mu_J}{\t}\right)\right]
        \\ 
	+ \left[\frac{\epsilon^i_{I}-v_k\beta^{ki}_{I}}{\t}- \nabla^i \left(\frac{\mu_I}{\t}\right)\right]
	\left[\tilde\s_{IJ} \t \frac{\epsilon_{Ji}-v^k\beta_{Jki}}{\t}
          - \tilde\xi_{IJ}\nabla_i
          \left(\frac{\mu_J}{\t}\right)\right] \\ 
	- \left[\frac{\epsilon^i_{I}-v_k\beta^{ki}_{I}}{\t} -
          \nabla^i\left(\frac{\mu_I}{\t}\right) \right]\tilde
        m_I\nabla_i p \\ 
	+ \left[\frac{\epsilon^i_{I}-v_k\beta^{ki}_{I}}{\t} - \nabla^i \left(\frac{\mu_I}{\t}\right) \right]\lbr
		\omega_I \Xi_i  
		- \tilde{\omega}_{IJ}\frac{2}{u^+}E^j_J \epsilon_{ij}
	\rbr,
\end{multline}
where $j_S^i$ is the non-relativistic canonical entropy current
defined as (similar to relativistic canonical entropy current given by
\eq{E:JmuSDef}), 
\be
	j^i_S = s v^i 
	- \frac{\mu_I}{\t} \vs^i_I.
\ee
%
We can rewrite \eq{E:entropyNRfull} in more conventional form as
\begin{multline}\label{E:entropyNRcon}
	\nabla_+ s
	+ \nabla_i j^i_S
	=
	\frac{1}{\t}\frac{n}{2}\sigma^{ij}\sigma_{ij}
	+ \frac{1}{\t}z(\nabla_k v^k)^2 \\
	+ \frac{1}{\t}\nabla^i\left[\kappa \nabla_i \t +
          \t\s_I\nabla^i\left(\frac{\mu_I}{\t}\right) -
          \s_I(\epsilon^i_{I}-v_k\beta^{ki}_{I})\right] \\ 
	+ \left[\frac{\epsilon_{Ji}-v^k\beta_{Jki}}{\t} - \nabla_i
          \left(\frac{\mu_J}{\t}\right)\right]\vs^i_I.
\end{multline}
The first three terms on the right hand side survive for uncharged
dissipative fluid \cite{landau}.  Here we compute the other terms
which are responsible for local entropy production for parity-odd
charged fluid in presence of background fields. Since we demand that
the fluid satisfies the second law of thermodynamics, the right
hand side of Eqn (\ref{E:entropyNRfull}) should be positive
definite. This implies following constraints on the transport
coefficients
\bea{ \label{E:NRconstraints1} 
\tilde m_I = 0, \quad n \geq 0, \quad z
  \geq 0,\quad
  \tilde\s_{IJ}=\frac{1}{\t}\tilde\xi_{IJ}, \nn \\
  \left[\tilde\xi_{IJ} - \frac{\xi_J q_I }{2(\epsilon+p)}\right] \quad
  \text{matrix is positive definite}.  } 
  
\begin{table}[t]
\caption{\label{tab:Inr1}%
Relations between relativistic parameters and non-relativistic
parameters (parity-even fluid)
}
\begin{ruledtabular}
\begin{tabular}{|c|c|c|}
\multicolumn{2}{|c|}{\textrm{Non-rel. variables}} &
\textrm{Rel. variables\footnote{$\mathbf{Z}$ is given by Eqn. (\ref{E:ZDef}), $\mathbf{Y}^i$ is given by Eqn. (\ref{E:YmuDef}) and $\chi$ is given by Eqn. (\ref{E:thermocorrec3}).}}\\
	\hline
	Velocity	&	$v^i$		&	$\displaystyle
        \frac{u^i}{u^+} - \frac{\eta}{(u^+)^2 (E+P)}\mathbf{Y}^i$ \\ 
	\hline
	Mass Density	&	$\rho$		&	$\displaystyle
        (u^+)^2(E+P) +(u^+)^2\left(\eta\mathbf{Z} -\zeta
          \theta\right)$ \\ 
	\hline
	Energy Density	&	$\epsilon$	&	$\displaystyle
        \frac{1}{2}(E-P) - \frac{1}{2}\left(\eta \mathbf{Z} - \zeta
          \theta \right)$ \\ 
	\hline
	Pressure	&	$p$		&	$\displaystyle
        P -\left(\frac{\eta}{d}\mathbf{Z}-\zeta \frac{u^\alpha
            \nabla_\alpha P}{E+P}	\right)$ \\ 
	\hline
	Charge		&	$q_I$		&	$\displaystyle
        u^+Q_{I}- u^+\mho_I\epsilon^{ij}u^+\nabla_i v_j$ \\ 
	\hline
	Entropy		&	$s$		&	$Su^+(1+\chi)$ \\
	\hline
	Scalar Potential	&	$\phi_I$		&	$A^-_I$ \\
	\hline
	Vector Potential	&	$a^i_I$		&	$A^i_I$
\end{tabular}
\end{ruledtabular}
\end{table}
These constraints are compatible with what we get from relativistic
entropy production condition. However the last term in
\eq{E:entropyNRfull} is a parity odd term and can not be written as a
perfect square. Therefore, it seems that the presence of this term
breaks the second law of thermodynamics. However, similar to the
relativistic fluid dynamics, we can add extra parity-odd terms
(allowed by symmetry) to entropy current to make right hand side a
perfect square.

Note that the last term in Eq. (\ref{E:entropyNRfull}) appears only in
$2+1$ dimensions.  Since the first order parity-odd terms in
\eq{E:entropyNRfull} do not appear in the non-relativistic fluid theories in more than two
spatial dimensions, we have a complete and consistent
description\footnote{This is the most generic description of a
  non-relativistic fluid obtained by light-cone reduction of a
  relativistic fluid system} for these theories. We summarize the results for them
in Tables \reff{tab:Inr1}, \reff{tab:Dr1}.

\begin{table}[t]
  \caption{\label{tab:Dr1}%
    Relations between relativistic and non-relativistic
    transport coefficients, energy momentum tensors and currents (parity-even fluid)
  }
\begin{ruledtabular}
\begin{tabular}{|c|c|c|}
\multicolumn{2}{|c|}{\textrm{Non-rel. variables}} &
\textrm{Rel. variables}\\
\hline
	\begin{minipage}{60px}Bulk Viscosity\end{minipage}	&
        $z$		&	$\zeta u^+$ \\ 
	\hline
	\begin{minipage}{60px}Shear Viscosity\end{minipage}	&
        $n$		&	$\eta u^+$ \\ 
	\hline 
	\begin{minipage}{60px} Electrical Conductivity\end{minipage} 	&	$\tilde\s_{IJ}$	&
        $\l_{IJ} u^+ + \eta u^+ Q_I Q_J/(E+P)^2$ \\ 
	\hline
	\begin{minipage}{60px}Thermal Conductivity\end{minipage}
        &	$\kappa$	&	$2n(\epsilon + p)/\t \rho$ \\ 
	\hline
			&	$\tilde\k_I$	&	$\k q_I/ 2(\epsilon + p)$ \\
	\hline
	\begin{minipage}{60px}Charge Conductivity\end{minipage}
        &	$\s_I$	&	$nq_I/\rho$ \\ 
	\hline
	\begin{minipage}{60px}Momentum Current\end{minipage}	&
        $t^{ij}$	&	$v^i v^j\rho+ p g^{ij}-n\sigma^{ij}-z
        g^{ij} \nabla_k v^k$ \\ 
	\hline
	\multirow{3}{*}{\begin{minipage}{60px}Energy
            Current\end{minipage}}	&	\multirow{3}{*}{$j^i$}
        &	$\dsp  v^i \left(\epsilon + p +\frac{1}{2}\rho \mathbf{v}^2\right)$ \\
		&	& $\dsp -n \sigma^{ik} v_k -zv^i\nabla_k v^k -\kappa\nabla^i \t$ \\
		&	& $\dsp - \t\s_I
                \nabla^i\left(\frac{\mu_I}{\t}\right)+\s_I
                \lb\epsilon^i_I - v_j\beta^{ji}_I\rb$ \\ 
	\hline
	\multirow{2}{*}{\begin{minipage}{60px}Charge
            Current\end{minipage}}	&
        \multirow{2}{*}{$j^i_I$}	&	$\displaystyle q_Iv^i
        -\tilde\k_I\nabla^i \t- \t\tilde\s_{IJ}\nabla^i\left(
          \frac{\mu_J}{\t}\right)$ \\ 
		&	& $\displaystyle +
                \tilde\s_{IJ}\lb\epsilon^i_{J} - v_k \beta^{ki}_{J}
                \rb$ \\ 
	\hline
	\begin{minipage}{60px}Entropy Current\end{minipage}	& $j^i_S$
	& \begin{minipage}{150px}
		$\displaystyle s v^i +
                \frac{\mu_I}{\t}\Big(\tilde\k_I\nabla^i t +
                \t\tilde\s_{IJ}\nabla^i\left( \frac{\mu_J}{\t}\right)$ 
		$\dsp - \tilde\s_{IJ}\lb\epsilon^i_{J} - v_k \beta^{ki}_{J} \rb\Big)$
	\end{minipage}
\end{tabular}
\end{ruledtabular}
\end{table}

\subsection{Parity-odd corrections to NR Entropy Current}

For fluids in two spatial dimensions however, we must add some
parity-odd terms to the entropy current with some arbitrary transport
coefficients and determine them from the condition that there is no
local entropy loss during the flow of the system. We make the most
generic parity-odd modification to the entropy current in two spatial
dimensions as follows
\bem{
	j^i_S \ra j^i_S
	+ \Big\{ \mathfrak{a}\epsilon^{ij}\nabla_j \t
	+ \mathfrak{b}_I\epsilon^{ij}\nabla_j\left(\frac{\mu_I}{\t}\right) 
	+ \mathfrak{c}\epsilon^{ij}\nabla_j p \\
	+ \mathfrak{d}_I\epsilon^{ij}\epsilon_{Ij} 
	+ \mathfrak{f}_I\epsilon^{ij}v^k\beta_{Ikj} \Big\}.
}
$\mathfrak{a}$, $\mathfrak{b}_I$, $\mathfrak{c}$, $\mathfrak{d}_I$ and
$\mathfrak{f}_I$ are arbitrary transport coefficients. After adding
these extra terms the right hand side of \eq{E:entropyNRfull} can be
  written as (parity-odd part only)
\bem{\label{E:NREntropyAnom}
	\Big\{\nabla_i\left(
		\mathfrak{a}\epsilon^{ij}\nabla_j \t
		+ \mathfrak{b}_I\epsilon^{ij}\nabla_j\left(\frac{\mu_I}{\t}\right) 
		+ \mathfrak{c}\epsilon^{ij}\nabla_j p
		+ \mathfrak{d}_I\epsilon^{ij}\epsilon_{Ij} \dbrk
		+ \mathfrak{f}_I\epsilon^{ij}v^k\beta_{Ikj}
	\right)\Big\}
	+ \left[ \frac{1}{T}E^i_{I} - \nabla^i \left(\frac{\mu_I}{\t}\right) \right]\lbr
		\omega_I \Xi_i 
		- \tilde{\omega}_{IJ}\frac{2}{u^+}E^j_J \epsilon_{ij}
	\rbr \\
	=
	\lbr\epsilon^{ij}\nabla_j \t\rbr\nabla_i\mathfrak{a}
	+ \lbr\epsilon^{ij}\nabla_j\left(\frac{\mu_I}{\t}\right)\rbr\nabla_i\mathfrak{b}_I
	+ \lbr\epsilon^{ij}\nabla_j p\rbr\nabla_i\mathfrak{c} \\
	+ \lbr\epsilon^{ij}\epsilon_{Ij}\rbr\nabla_i\mathfrak{d}_I
	+ \lbr\half v^k\epsilon^{ij}\beta_{Iij}\rbr\nabla_k\mathfrak{f}_I \\
	- \lbr \epsilon^{ij}\nabla_i \left(\frac{\mu_I}{\t}\right)\rbr\left(
		\frac{\omega_I}{\t (u^+)^2}\nabla_j \t +
                \omega_I\frac{q_I \t}{\rho}
                \nabla_j\left(\frac{\mu_I}{\t}\right) -
                \frac{2\omega_I}{\rho}\nabla_j p 
	\right) \\
	+ \lbr \epsilon^{ij}\frac{1}{T}E_{Ii}\rbr\left(
		\frac{\omega_I}{\t (u^+)^2}\nabla_j \t \dbrk
		+ \frac{\t}{\rho}\left(\omega_Iq_J + \omega_Jq_I -
                  2\rho\tilde{\omega}_{JI}\right)
                \nabla_j\left(\frac{\mu_J}{\t}\right)  
		- \frac{2\omega_I}{\rho}\nabla_j p
	\right) \\
	+ \lbr \epsilon^{ij}\frac{1}{T}E_{Ii}\rbr\left(
		\frac{\omega_I}{\rho}Q_J E_{Jj}
		- \tilde{\omega}_{IJ}\frac{2}{u^+}E_{Jj}
	\right) \\
	+ \half\lbr\mathfrak{f}_I\epsilon^{ij}\beta_{Iij}\nabla_k v^k
	- \mathfrak{d}_I\epsilon^{ij}\nabla_+\beta_{Iij}\rbr
	= 0.
}
To obey the second law of thermodynamics we claim that the
coefficients of all the independent terms must vanish.  This is
possible except for the last two terms, for which we get the following
sets of constraints:
\bea{
	\frac{\partial \mathfrak{a}}{\partial p}
	-\frac{\partial \mathfrak{c}}{\partial \t} &= 0, \nn\\
	\frac{\partial \mathfrak{a}}{\partial (\mu_I /\t)}
	- \frac{\partial \mathfrak{b}_I}{\partial \t}
	&= \frac{\epsilon + p}{\t}\frac{2\omega_I}{\rho},  \nn\\
	\frac{\partial \mathfrak{b}_I}{\partial p}
	- \frac{\partial \mathfrak{c}}{\partial (\mu_I /\t)}
	&= \frac{2\omega_I}{\rho}, \label{E:anomalyDE3}\\
	- \frac{\partial \mathfrak{f}_I}{\partial \t}
	= \frac{\partial \mathfrak{d}_I}{\partial \t} 
	&= \frac{\epsilon + p}{\t^2}\frac{2\omega_I}{\rho}, \nn\\
	\frac{\partial \mathfrak{f}_I}{\partial p}
	= - \frac{\partial \mathfrak{d}_I}{\partial p}
	&= \frac{2\omega_I}{\t\rho}, \nn\\
	- \frac{\partial \mathfrak{f}_I}{\partial (\mu_J /\t)}
	= \frac{\partial \mathfrak{d}_I}{\partial (\mu_J /\t)}
	&= \frac{1}{\rho}\left(\omega_Iq_J + \omega_Jq_I -
          2\rho\tilde{\omega}_{JI}\right), \label{E:anomalyDE6} 
}
additionally, the following matrices must be symmetric:
\bea{
	2\rho\tilde{\omega}_{IJ}
	- \omega_Iq_J, \nn\\
	\frac{\rho}{t}\frac{\partial \mathfrak{b}_J}{\partial (\mu_I /t)}
	- \omega_Iq_J. \label{E:anomalyDE8}
}

It should be noted that the new coefficients introduced in $j^i_S$
need not be most generic, and their only significance is to make the
parity-odd terms in entropy current vanish. Hence, a minimal choice is
as follows:
\bee{
	\fa = \fc = 0, \qquad \ff_I = - \fd_I,
}
which solves the respective constraints, and leaves us only two
independent coefficients in entropy current: $\fb_I$ and $\fd_I$.

Now we turn attention towards the last two terms of
\eq{E:NREntropyAnom}. They vanish if we consider the fluid is
incompressible and placed in a time independent background. For
compressible fluid the second last term is non-zero and can be
negative. Hence the second law implies that the $\ff_I = \fd_I =0$. As a
result all the parity-odd terms vanish for compressible fluid. 
The resultant fluid is completely parity-even and we summarize the results in tables \reff{tab:Inr1},
\reff{tab:Dr1}. On the
other hand if we consider the background field is time dependent, then
the last term can violate the second law. However, in that case, one
has to be careful about the change of entropy of the background
also. The second law should hold for the fluid and the background
system together. The last term produces some vortex kind of motion in
the fluid. In that case the velocity of fluid particles around those
vortexes becomes very large and our analysis (derivative expansion)
may break down.



It should be noted that we demand our non-relativistic system to obey
the second law of thermodynamics. We start with a relativistic fluid
but never impose any physicality constraint (except the fluid
satisfies first law of thermodynamics) on the relativistic
side. Relativistic system, in our case, can be considered as a
mathematical model (where we allow all possible terms allowed by the
symmetry). After reduction we see that at least for the parity-even
sector constraint from the second law turns out to be same
(compatible) both in relativistic and non-relativistic case
(Eq. \eq{E:Rconstraints1} and \reff{E:NRconstraints1}). But for
parity-odd sectors they are completely different. For example,
$C_{IJK}$ do not even appear in non-relativistic relations
\eq{E:anomalyDE3} - \reff{E:anomalyDE8}, though for relativistic fluid
the second law relates $\mho_I$ and $\tilde{\mho}_{IJ}$ to $C_{IJK}$
(Eq. \eq{E:anomalyRelDE}). Whereas for non-relativistic fluid one can
consider $\o_I$ and $\tilde \o_{IJ}$ to be independent and $\ff_I$ and
$\fd_i$ are fixed in terms of them. However, one can also start with a
relativistic fluid satisfying second law of thermodynamics (as an
additional condition). In that case for an incompressible fluid it is
possible to express the non-relativistic transport coefficients
($\ff_I, \fd_I, \omega_I, \tilde \o_I$) in terms of anomaly
coefficients of the relativistic system. But for compressible fluid it
is not possible as these coefficients turn out to be zero

\subsection{Incompressible fluid in constant magnetic background in
  (2+1)-dim}\label{incomp} 

Equation of state of an incompressible fluid is given by: 
\bee{ \r =
  \text{constant}, }
which implies that $\displaystyle \nabla_k v^k = 0$, or in other
words, there is no compression or expansion of the fluid during the
flow. 

Using the non-relativistic constitutive equations one can easily show that:
\bee{
	\frac{\df \epsilon}{\df t} + (\e + p)\nabla_k v^k = v_j \nabla_i \pi^{ij} - \nabla_i \vs^i,
}
which means that for incompressible fluids, ${\df \epsilon}/{\df t}$
is atmost two derivative in order, and is zero for ideal
fluids. Considering this result at first derivative order we can infer
that:
\bee{
	u^\mu\nabla_\mu (E-P) = 0 \Ra \frac{u^\mu\nabla_\mu P}{E+P} = -\theta,
}
in first derivative order. Same can be inferred directly from \eq{E:thetadvkrelation}. \\

Maxwell's equation in the magnetic limit \apndxIn{nrelectro} are
given by:
\bea{
	\nabla_i \epsilon^i_I &= 0, \\
	\nabla_i \beta^{ij}_I &= 0, \\
	2\epsilon^{ij}\nabla_i \epsilon_{Ij}+\epsilon^{ij}\nabla_+\beta_{Iij} &=0.
}
Second equation implies that magnetic field is constant in
space. Also, if magnetic field $\beta^{12}_I$ is time independent (as
we demand), electric field is curl free. But first equation already
tells us that electric field should be divergence-free. Hence electric
field is constant over the 2 dimensional space. However, electric
field can still be time dependent, as the corresponding term does not
appear in the magnetic limit of Maxwell's equations. Finally in
our case we have, a constant electric field (may depend on time) and a
constant magnetic field.

For this system, the form of $j^i_S$ is given by
\bem{
	j^i_S = s v^i 
	- \frac{\mu_I}{\t} \upsilon^i_I
	+ \mathfrak{b}_I\epsilon^{ij}\nabla_j\left(\frac{\mu_I}{\t}\right) 
	+ \mathfrak{d}_I\epsilon^{ij}\left(\epsilon_{Ij} - v^k\beta_{Ikj}\right).
}
where $\mathfrak{b}_I$ and $\mathfrak{d}_I$ are determined by
equations \reff{E:anomalyDE3} - \reff{E:anomalyDE8}. Hence we will get
the positivity of entropy in the form of \eq{E:entropyNRcon},
along with constraints \eq{E:NRconstraints1}. However $\omega_I$ and
$\tilde{\omega}_{IJ}$ remain unconstrained by entropy
positivity. Results of this case have been summarized in Tables
\reff{tab:Dr2}.


\begin{table}[t]
  \caption{\label{tab:Dr2}%
    Relativistic and non-relativistic transport coefficients for
    parity-odd fluid
  }
\begin{ruledtabular}
\begin{tabular}{|c|c|c|}
\multicolumn{2}{|c|}{\textrm{Non-rel. variables}} &
\textrm{Rel. variables}\\
\hline
	Parity-odd	&	$\omega_I$	&	$\mho_I
        (u^+)^2$ \\
coefficients
				&	$\tilde{\omega}_{IJ}$	&	$\tilde{\mho}_{IJ} u^+$ \\
	\hline
	Mom. Curr.	&	$t^{ij}$	&	$v^i v^j\rho+ p g^{ij}-n\sigma^{ij}$ \\
	\hline
	Energy	&	\multirow{2}{*}{$j^i$}		&
        $\displaystyle  v^i \left(\epsilon + p +\frac{1}{2}\rho
          \mathbf{v}^2\right)-n \sigma^{ik} v_k$ \\ 
	Current	&	& $\displaystyle -\kappa\nabla^i \t - \t\s_I
        \nabla^i\left(\frac{\mu_I}{\t}\right)+\s_I \lb\epsilon^i_I -
        v_j\beta^{ji}_I\rb$ \\ 
	\hline
		&	\multirow{4}{*}{$j^i_I$}	&
                $\displaystyle q v^i - \tilde\k_I\N^i \t -
                \t\tilde\s_{IJ} \N^i\lb\frac{\mu_J}{\t}\rb$ \\ 
	Charge	&	& $\displaystyle  + \tilde\s_{IJ}\lb\epsilon^i_{J} - v_k \beta^{ki}_J \rb$ \\
	Current	&	& $\displaystyle + \o_I
        \e^{ij}\Big(\frac{\k}{n}\N_j \t  + \frac{\s_J \t}{n}
        \N_j\lb\frac{\mu_J}{\t}\rb 
	-\frac{2}{\r} \N_j p \Big)$ \\
		&	& $\displaystyle +
                \bar\s_{IJ}\e^{ij}\lb\epsilon_{Jj} - v^k \beta_{Jkj}
                \rb$ \\ 
	\hline
		&	\multirow{5}{*}{$j^i_S$}	&
                $\displaystyle s v^i +
                \frac{\mu_I}{\t}\Big(\tilde\k_I\N^i \t +
                \t\tilde\s_{IJ} \N^i\lb\frac{\mu_J}{\t}\rb$ \\ 
	Entropy	&	& $\displaystyle -
        \tilde\s_{IJ}\lb\epsilon^i_{J} - v_k \beta^{ki}_J \rb\Big)$ \\ 
	Current	&	& $\displaystyle - \frac{\mu_I}{\t}\o_I
        \e^{ij}\Big(\frac{\k}{n}\N_j \t -\frac{2}{\r} \N_j p \Big)$ \\ 
		&	& $\displaystyle + \lb \mathfrak{b}_J -
                \frac{\o_I\mu_Iq_J}{\r}\rb\e^{ij}
                \N_j\lb\frac{\mu_J}{\t}\rb$ \\ 
		&	& $\displaystyle + \lb \mathfrak{d}_J -
                \frac{\mu_I}{\t}\bar\s_{IJ}\rb \e^{ij}\lb\epsilon_{Jj}
                - v^k \beta_{Jkj} \rb$ 
\end{tabular}
\end{ruledtabular}
\end{table}

\section{Discussion}

We started with a generic $(d+2)$-dim non-conformal relativistic fluid
($d\geq 2$) with anomalies specific to $d=2$ in presence of background
electromagnetic fields on flat space. Light cone reduction of this system gives a
$(d+1)$-dim non-relativistic fluid. We demand that our
non-relativistic fluid satisfies the second law of thermodynamics
which imposes certain constraints on the system. For example we find
that the parity odd terms in non-relativistic theory can only sustain
for incompressible fluid in electromagnetic background with constant
magnetic field in $(2+1)$-dim.

Our non-relativistic fluid (obtained by LCR) has generic
dissipative terms, which are allowed by the symmetry and the condition
that local entropy production is always positive definite. LCR does
not constraint the size of these coefficients. They all are fixed in
terms of the transport coefficients of the `mother' relativistic
theory. On the contrary, when one performs a $1/c$ expansion the
relativistic constitutive equations to get a non-relativistic fluid
system, as done in \cite{Kaminski:2013gca}, many of these terms are
suppressed depending on the physical considerations and the type of
system under view. We discuss the basic aspects of $1/c$ expansion in
\apndx{1bycapnda}.

Apart from the dissipative terms, our system and the system obtained
by $1/c$ expansion in \cite{Kaminski:2013gca} have certain fundamental
differences. Firstly, their system is `extensive' as the thermodynamic
variables follow Euler's relation; however since LCR breaks the
Euler's relation, our system is no longer extensive. Secondly, in our
system $\r$ is not necessarily proportional to $q_I$, while in
\cite{Kaminski:2013gca} it is true at least at the leading $1/c$
order. This is why our non-relativistic system has one more
independent parameter as opposed to the $1/c$ case. We can however
enforce $\r \propto q_I$ in our system as well, but it turns out that
demanding so switches off all the dissipative terms from the theory
except for the bulk viscosity.

In \cite{Kaminski:2013gca} authors do not present an entropy current
calculation for non-relativistic fluid obtained by $1/c$ expansion. In
fact, as we will review in \apndx{1bycapnda}, the entropy positivity
turns out to be trivial, and is just followed from the leading order
entropy current of the relativistic theory. The constraints on the
transport coefficients (which survives at the leading order) also
turns out to be the same. However in our case, we have slightly
different constraints, because of the above mentioned fundamental
differences between the two cases.\\


\noindent
{\bf Acknowledgement}\\

We would like to acknowledge Pratik Roy and Rahul Soni for valuable
discussion. SD would like to thank the hospitality of $IACS, \
Kolkata$ where part of this work has been done. ND would like to thank
the DST $Ramanujan$ Fellowship. Finally we are indebted to the people
of India for their unconditional support towards research and
development in basic science.

\appendix

\section{Magnetic limit of electrodynamics}\label{nrelectro}

Maxwell's Electrodynamics is a relativistic theory. In fact it was a precursor to Einstein's Special Theory of Relativity. Having a consistent relativistic description of electrodynamics, eradicated any need for a `non-relativistic' theory of electrodynamics. However as in current context, one needs a Galilean description of electrodynamics, just to keep it consistent with other non-relativistic theories.

Recently in \cite{Manfredi:2013}, authors discussed the non-relativistic limit of electrodynamics in two distinct ways. Depending on the strength of fields\footnote{In this section we use the conventional notation for electrodynamics: $E$ for electric field, $B$ for magnetic field, $\r$ for charge density, $J$ for charge current etc.}, we can have two discrete non-relativistic limits: $|\vec E| \gg c |\vec B|$ electric limit and $|\vec E| \ll c |\vec B|$ magnetic limit. We find that the light-cone reduction of relativistic sourceless Maxwell's equations, under certain identifications, gives magnetic limit of Maxwell's equations, which will be our interest of discussion here. A more thorough discussion of non-relativistic electrodynamics can be found in \cite{Manfredi:2013}.

In arbitrary dimensions, inhomogeneous Maxwell's equations are given by:
\bea{
	\N_i F^{i0} = - \mu_o c \r, &\qquad
	\N_0 F^{0i} + \N_j F^{ji} = - \mu_o J^i,
}
while the homogeneous ones (Bianchi identities) are:
\bee{ \label{E:bianchi}
	\e^{\a\b\cdots\mu\nu\s}\N_{\mu}F_{\nu\s}= 0,
}
where $\e^{\a\b\cdots\mu\nu\s}$ is the full-rank Levi-Cevita tensor. The last identity follows directly from the gauge invariant form of $F^{\mu\nu} = \N^\mu A^\nu - \N^\nu A^\mu$ or the constituent fields\footnote{Instead of the conventional magnetic field, we use its 2nd rank dual in our work, as in arbitrary dimensions we do not have electromagnetic duality and magnetic field does not have any fixed rank, while its dual has.}: $E^i = -\N^i \f - \N_0 A^i$ and $F^{ij} = \N^i A^j - \N^j A^i$. As LCR gives a non-relativistic theory with gauge invariance, we are interested in a non-relativistic limit which preserves \eq{E:bianchi}. We will see that the magnetic limit essentially does the same. However there exists a consistent electric limit as well, where Bianchi identities are broken, but inhomogeneous Maxwell's equations are preserved.

In terms of dimensionless fields and sources one can write the inhomogeneous Maxwell's equations as:
\bea{\label{E:redmaxwell}
	\a\N_i E^{i} = \r, &\qquad
	\b\frac{\dow}{\dow t} E^{i} + \N_j F^{ji} = - \frac{\b}{\a} J^i,
}
and the Bianchi identities as:
\bee{
	\e_{lm\cdots ij}\N^{i}E^{j} = - \b\half \e_{lm\cdots ij}\frac{\dow}{\dow t}F^{ij},
}
\bee{
	\e^{m\cdots k ij}\N_{k}F_{ij} = 0,
}
where $\a$ and $\b$ are some dimensionless constants given by:
\bee{
	\a = \frac{[E]\e_o}{[\r][L]}, \qquad \b = \frac{[L]}{[T]c}.
}
From here we can easily read out, that in non-relativistic limit $\b\ll 0$. To preserve the Bianchi identities therefore, $E^i$ should be one order smaller than $F^{ij}$; which is why this limit is called `magnetic limit'. If we measure smallness in terms of a parameter $\e \sim 1/c$, we will have $E^i \sim \e^{n+1}$ and $F^{ij} \sim \e^{n}$. Therefore from \eq{E:redmaxwell}, first nontrivial order of $\r\sim \a\e^{n+1}$, and $J^i \sim \a \e^{n-1}$, and inhomogeneous Maxwell's Equations (in conventional units) reduce to:
\bea{\label{E:magmaxwell}
	\N_i E^{i} = \frac{\r}{\e_o}, &\qquad
	\N_j F^{ji} = - \mu_o J^i.
}
Essentially we have just dropped the displacement current term from the Ampere's Law. To explicitly see if this limit is Galilean invariant, one can lookup \cite{Manfredi:2013}. These are the very equations that have been used in \sctn{rednofback} and \reff{incomp}.


One can now push the expansion of \eq{E:redmaxwell} to $\e^{n+2}$ order and derive the continuity equation:
\bee{
	\frac{\dow}{\dow t} \r + \nabla_i J^i = 0.
}
It should be remembered however that there are mixed orders in this equation and to the highest order it just states: $\nabla_i J^i = 0$, which also follows from \eq{E:magmaxwell}. Continuity equation takes its usual form only when first two leading orders of $J^i$ vanish and $\r \sim J^i$, but in which case one of the Maxwell's equations modifies to: $\N_j F^{ji} = 0$. This in conjugation with the Bianchi identity would mean that the magnetic field is constant over space.

Finally in conventional units this limit can be summarized as: $F^{ij} \sim E^i \sim c^{-n}$, and to the maximum order $\r \sim c^{-n}\e_o$, $J^i \sim c^{-n+2}\e_o$.

\section{1/c expansion of relativistic fluid dynamics} \label{1bycapnda}

In this section we discuss briefly the $1/c$ expansion limit of relativistic fluid dynamics to get a non-relativistic theory, using the prescription of \cite{Kaminski:2013gca}. We consider here just a parity-even fluid for comparison with our work. Parity-odd sector in our case and in \cite{Kaminski:2013gca} are in different hydrodynamic frames and thus are incomparable.

The constitutive equations of a relativistic fluid (with appropriate factors of $c$) are:
\bee{
	\N_\mu T^{\mu\nu}	= c F^{\nu\a}_I J_{I\a}, \qquad
	\N_\mu J^{a\mu}_I	= 0,
}
where
\bee{
	T^{\mu\nu}	= (E+P) u^\mu u^\nu + P g^{\mu\nu} + \Pi^{\mu\nu},
}
\bee{
	J^{a\mu}_I	= Q_I u^\mu + \U^{\mu}_I.
}
Respective dissipative terms are given as:
\bee{
	\Pi^{\mu\nu} = -2 \eta \tau^{\mu\nu} - \zeta \theta P^{\mu\nu},
}
\bee{
	\Upsilon^{\mu}_{I} =
	- \frac{1}{c^2}T\l_{IJ}P^{\mu\nu}\nabla_\nu \left(\frac{M_J}{T}\right) 
	+ \frac{1}{c}\lambda_{IJ} E^\mu_J.
}
We can separate out rest contributions from $E$ and $M_I$:
\bee{
	E	= R	c^2 + \cE, \qquad
	M_I	= \frac{1}{\cK}m_I	c^2 + \mu_I.
}
Here $m_I$ is the mass is to `$I$'th charge ratio of constituent particles in their local rest frame, which is assumed to be constant. $\cK$ is the total number of $U(1)$ charges. Non-relativistic mass and energy density are related to their relativistic counterparts as:
\bee{
	\r	= R\G, \qquad
	\e	= \cE\G,
}
where $\G = (1-\mathbf{v}^2/c^2)^{-1}$. Since the fluid under consideration is single component, $R/Q_I = m_I$ is a constant. Non-relativistic charge density is defined as:
\bee{
	q_I	= J_I^0.
}
Pressure and Temperature are however kept same, which to be consistent with the main text notation we will denote as: $\t = T$ and $p = P$.

Before continuing with expansion we need to fix the order of various quantities. $\r,\e, p, \t, v^i$ can be thought of finite order without much ambiguity. $m_I$ on the other hand is quite sensitive to the kind of system under consideration. Let us consider a charged fluid made of `ions' where $m_I \sim c^2$; $m_I$ cannot be too low, or else the fluid would start coupling to the background fields. Correspondingly the charge density $q_I$ would be of order $c^{-2}$. Further, to keep the thermodynamics intact, one has to assume $\mu_I$ to be of order of $c^2$. Finally, for external fields to have finite effect (e.g. force) on the fluid, $\e^i_I \sim \b^{ij}_I \sim c^2$.

Using this information we can reduce the energy-momentum tensor to:
\bea{
	T^{00} &=  \r c^2 + \half \r \mathbf{v}^2 + \e + \cO(1/c), \nn \\
	T^{i0} &= \r v^i c + \lb\half \r \mathbf{v^2} + \e +p\rb \frac{v^i}{c} + \frac{1}{c}\pi^{ij}v_j + \cO(1/c^2), \nn \\
	T^{ij} &= t^{ij} + \cO(1/c), \qquad t^{ij} = \r v^iv^j + p g^{ij} + \pi^{ij},
}
where we have used:
\bea{
	n = \frac{\eta}{c}, \qquad z = \frac{\zeta}{c},
}
\bee{
	\pi^{ij} = - n \lb\N^i v^j + \N^j v^i - g^{ij}\frac{2}{d} \N_k v^k\rb - z g^{ij} \nabla_k v^k.
}
Energy-momentum conservation equations, at highest order, will then reduce to:
\bee{
	\dow_t \r + \dow_i (\r v^i) = 0,
}
\bee{
	\dow_t (\r v^i) + \dow_i t^{ij} = \e^{i}_I q_I + \b^{ij}_I j_{Ij}.
}
It will be worth to mention here the underlying assumption: $n,z \sim 1$ which is just an empirical fact. This is precisely the reason why no dissipative corrections appear to the continuity equation. We can now expand the charge current:
\bea{
	J^{0}_I	&= q_I, \qquad q_I = Q_I \G + \cO(1/c^6), \nn \\
	J^{i}_I	&= \frac{1}{c} j^{i}_I + \cO(1/c^7), \qquad j^{i}_I = q_I v^i + \vs^{i}_I,
}
where we have used:
\bee{
	\tilde\s_{IJ} = \frac{\l_{IJ}}{c}, \qquad \tilde\g_{IJ} = \frac{\g_{IJ}}{c},
}
\bee{
	\vs_I^{i} = c^2 \tilde\s_{IJ} \frac{m_J}{\cK\t} \nabla^i \t.
}
Here again we have used a physical input to fix the order of $\tilde\s_{IJ}$. We would demand that in the non-relativistic theory, charge to mass ratio should be a constant. Therefore charge continuity and mass continuity equations should be the same upto leading order in $1/c$, and hence $\vs_I^{i}$ should at maximum be of the order of $c^{-4}$. It further implies that $\tilde\s_{IJ} \sim c^{-8}$. As a consequence, effects like electric conductivity are suppressed in whole theory.

Finally we use: $\nabla_\mu \lb T^{\mu 0} - \frac{1}{\cK}m_I c^2 J^{\mu}_I \rb = 0$ to get the energy conservation:
\bee{
	\dow_t \lB\half \r \mathbf{v}^2 + \e \rB + \dow _i \lB v^i \lb \half \r \mathbf{v^2} + \e +P \rb + \vs^i \rB = 0,
}
\bee{
	\vs^i = \pi^{ij} v_j - \k \nabla^i \t,
}
where we have identified the thermal conductivity:
\bee{
	\k = \tilde\s_{IJ}\frac{c^4}{\cK^2}\frac{m^a_I m_J^a}{\t}.
}
which is of finite order.

The non-relativistic theory has essentially turned out to be identical to the chargeless fluid discussed in \cite{landau}, because for uni-component fluids, charge density serves just as number density upto a factor. For multi-particle fluids however this might become more interesting, as then one does not have to expect charge is to mass ratio to be constant. In fact one can have some chargeless particles also in the system.

\subsection{1/c Expansion of Thermodynamics}

First law of thermodynamics and the Euler's relation in the relativistic theory are given by:
\bee{
	\df E = T \df S + M_I \df Q_I,
}
\bee{
	E + P = TS + M_I Q_I.
}
Under $1/c$ expansion they reduce to:
\bee{
	\df \e = \e \df s + \mu_I \df q_I,
}
\bee{
	\e + p = \t s + \mu_I q_I,
}
which are just the non-relativistic analogues of the same equations. We therefore conclude that the non-relativistic system gained by $1/c$ expansion follows extensivity.

Now lets have a look at the second law of thermodynamics in the relativistic side, which mentions: $\N_\mu J^\mu_S \geq 0$. From \eq{E:entropyRel1} we know the positive definite form of entropy current:
\bem{
	\nabla_\mu J^\mu_S =
        \frac{1}{T}\eta\tau^{\mu\nu}\tau_{\mu\nu}+\frac{1}{T}\zeta
        \theta^2 \\
+ \left[\frac{E^\alpha_{I}}{c^2T}-\frac{1}{c}P^{\alpha\mu}\nabla_\mu
          \left(\frac{M_I}{T}\right)\right] \varrho_{IJ}
        \left[\frac{E_{J\alpha}}{c^2T}-\frac{1}{c}P_{\alpha\nu}\nabla^\nu
          \left(\frac{M_J}{T}\right) \right].
}
In highest order this equation for non-relativistic systems says that:
\bea{
	\N_t s + \nabla_i j^i_S =
        \frac{1}{2\t}n\s^{ij}\s_{ij}+\frac{1}{\t} z \lb \N_k v^k \rb
	+ \frac{\k}{\t^2}\lb\nabla^i \t\rb^2.
}
But if we look at the respective coefficients from the relativistic side:
\bee{
	n = \frac{\eta}{c} \geq 0, \qquad z = \frac{\zeta}{c} \geq 0,
}
\bee{
	\k = \varrho_{IJ}\frac{c^3}{\cK^2}\frac{m^a_I m_J^a}{T^2} \geq 0.
}
We hence see that the relativistic entropy positivity implies non-relativistic entropy positivity.

\end{document}